\documentclass[aps,prl,reprint,superscriptaddress,floatfix]{revtex4-2}
\usepackage{bm, color}
\usepackage{placeins}
\usepackage{nameref}
\usepackage{amssymb}
\usepackage{graphicx}
\usepackage{subfigure}
\usepackage{dcolumn}
\usepackage{bm}
\usepackage{feynmf}
\usepackage{mathtools, nccmath}
\usepackage{tikz}
\allowdisplaybreaks
\usepackage[colorlinks = true,
            allcolors  = blue]{hyperref}
\usepackage{color,soul}

\def\H{\mathcal{H}}

\usepackage{CJK}
\usepackage{hyperref}

\renewcommand{\eqref}[1]{Eq.~(\ref{#1})}

\preprint{APS/123-QED}
\begin{document}


\title{Global approximations to correlation functions of strongly interacting quantum field theories}
\author{Yuanran Zhu}%
\thanks{Corresponding author}
\email{yzhu4@lbl.gov}
\affiliation{Applied Mathematics and Computational Research Division, Lawrence Berkeley National Laboratory, Berkeley, USA, 94720.}

\author{Yang Yu}
\thanks{Corresponding author}
\email{umyangyu@umich.edu}
\affiliation{Department of Physics, University of Michigan, Ann Arbor, Michigan 48109, USA}

\author{Efekan K\"okc\"u}%
\affiliation{Applied Mathematics and Computational Research Division, Lawrence Berkeley National Laboratory, Berkeley, USA, 94720.}



\author{Emanuel Gull}
\affiliation{Department of Physics, University of Michigan, Ann Arbor, Michigan 48109, USA}
\affiliation{Department of Physics, University of Warsaw, Warsaw, Poland}




\author{Chao Yang}%
\affiliation{Applied Mathematics and Computational Research Division, Lawrence Berkeley National Laboratory, Berkeley, USA, 94720.}






\begin{abstract}
We introduce a method for constructing global approximations to correlation functions of strongly interacting quantum field theories, starting from perturbative results. The key idea is to employ interpolation method, such as the two-point Pad\'e expansion, to interpolate the weak and strong coupling expansions of correlation function. We benchmark this many-body interpolation approach on two prototypical models: the lattice $\phi^4$ field theory and the 2D Hubbard model. For the $\phi^4$ theory, the resulting two-point Pad\'e approximants exhibit uniform and global convergence to the exact correlation function. For the Hubbard model, we show that even at second order, the Pad\'e approximant already provides reasonable characterizations of the Matsubara Green’s function for a wide range of parameters. Finally, we offer a heuristic explanation for these convergence properties based on analytic function theory.
\end{abstract}

\maketitle
Developing accurate and numerically efficient computational methods for strongly interactive systems remains a central challenge in  quantum field theory. 
Non-perturbative numerical approaches, among them quantum Monte Carlo (QMC)~\cite{blankenbecler1981monte,zhang1995constrained,rubtsov2005continuous,Prokofev1998QMC,ceperley1995path,creutz1979monte,Gull11RMP}, 
tensor-network methods~\cite{white1992DMRG,schollwock2005density,jordan2008classical}, 
and dynamical mean-field theories~\cite{metzner1989DMFT,georges1992DMFT,georges1996dynamical,kotliar2006electronic,lichtenstein2000cluster,maier2005quantum} have provided invaluable  insights into the physics of the strong correlation regime of interactive field systems, including the notoriously subtle two-dimensional Hubbard model \cite{LeBlanc15,Arovas22,Qin22}. On the other hand, perturbative approaches \cite{MP2,Hedin65}, despite their limited accuracy in the strong correlation regime, remain attractive due to their analytic or semi-analytic nature, general applicability to electronic structure systems, and favorable scalability to large system sizes. Especially in recent years, 
several algorithmic developments in continuous-time quantum Monte Carlo \cite{Gull11RMP,rubtsov2005continuous,Gull08_ctaux,Werner06,Werner06Kondo}, diagrammatic Monte Carlo methods \cite{Prokofev1998QMC,Van_Houcke_2010,rossi2017determinant,Moutenet_CDet_2018}, and automatic diagram generations \cite{kleinert2000recursive,serone2018lambdaphi4,hahn2001generating} have made it possible to evaluate the perturbation series to relatively high orders for interactive quantum fields that appear in both condensed matter and high-energy physics. 
With these advances, the central challenge of solving strongly interative systems with perturbative methods has gradually shifted from their computational cost to the fundamental limitations imposed by the finite radius of convergence of the perturbation series itself.

Extending the applicability of diagrammatic expansions of quantum field theories beyond the convergence radius of the original perturbation series typically requires resummation techniques. Several approaches have been proposed, including Pad\'e approximants~\cite{baker1981pade,vsimkovic2019determinant,ferrero2023diagrammatic,serone2018lambdaphi4}, 
resurgence theory~\cite{dorigoni2019introduction,basar2013resurgence,Rossi_resummation_2018}, 
and action deformation strategies~\cite{Pollet_regularization_2010,Rossi_action_2016,Kim_mapping_2021}. In this paper, we propose an alternative approach based on a classical yet powerful idea from approximation theory -- the two-point Pad\'e expansion~\cite{bender2013advanced}. 
The general procedure is as follows: for an interacting field theory, we construct two series expansions for the correlation function, namely a weak-coupling expansion (WCE) and a strong-coupling expansion (SCE), which are valid in the limits of interaction strength \( g \to 0 \) and \( g \to +\infty \), respectively. We then apply multi-point interpolation methods, such as the two-point Pad\'e expansion, to interpolate between these two limits. This procedure naturally results in a global approximation to the correlation function for all $g\geq 0$. 

\begin{figure}[t]
\centering
\includegraphics[width=8.5cm]{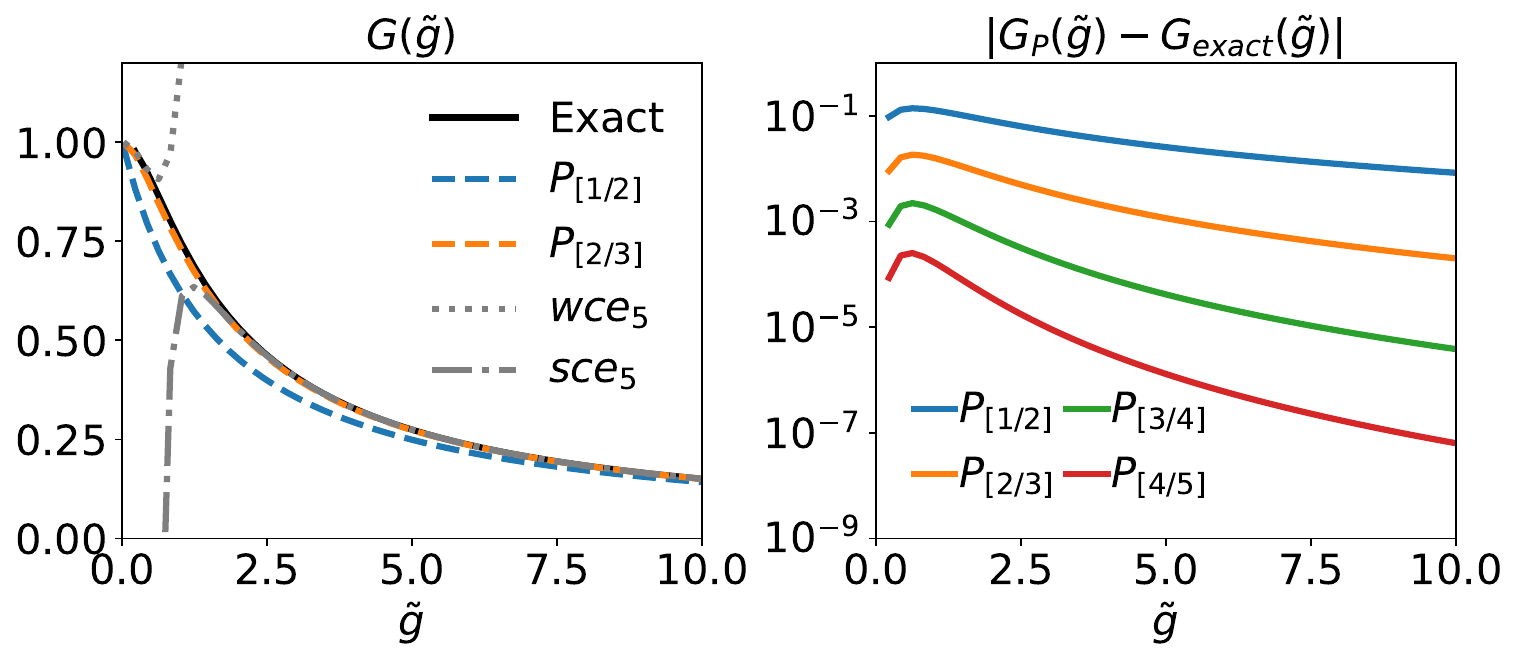}
\caption{\textbf{(Left)} Two-point Pad\'e approximations of \( G(\tilde{g}) \) for the \(0d\)\,-\,\(\phi^4\) model, compared with 5-th order WCE \eqref{0d_wce} and 5-th order SCE \eqref{0d_sce} expansion results. \textbf{(Right)} Approximation error of two-point Pad\'e expansions relative to the exact result \eqref{od_G_analytic}.
}
\label{fig:Gii_log} 

\end{figure}

While technically similar to the conventional \emph{one-point} Pad\'e expansion~\cite{baker1981pade, vsimkovic2019determinant, ferrero2023diagrammatic,serone2018lambdaphi4} used for WCE resummation, the proposed \emph{two-point} approach differs fundamentally in two aspects. First, incorporating information from both endpoints effectively transforms the \emph{extrapolation} problem into an \emph{interpolation} problem, which enables the construction of an approximant that converges {\em globally} for all interaction strengths \( g \ge 0 \). This property is in sharp contrast to the one-point method whose convergence is inherently restricted to regimes near the origin. We numerically demonstrate this convergence in the lattice $\phi^{4}$ theory and the Hubbard model and provide a heuristic justification for the observed convergence behavior.  Second, since the \emph{two-point} scheme matches two limits instead of one limit, a given order of Pad\'e approximation requires roughly half the perturbation series expansion order compared with the \emph{one-point} scheme. This dramatically reduces the computational cost arising from enumerating Feynman diagrams. We note that the idea of interpolating between weak-
and strong-coupling expansions has previously been applied to quantum field and string theory models in
high-energy physics~\cite{sen2013s,banks2013two,pius2014s} but, to our knowledge, these applications heavily rely on duality principle to get the SCE {\em from} the WCE. A key difference here is that we employ techniques to {\em directly} obtain the strong coupling expansion of lattice field theories, which enables us to address condensed matter systems such as the Hubbard model.  
\paragraph{A simple example.} We illustrate the central ideas using a
schematic model---the zero-dimensional $\phi^4$ model \cite{brown2015two}. Despite its simplicity, this example demonstrates how the two-point Pad\'e expansion interpolates the weak and strong coupling expansion for a general field theory model and yields {\em globally convergent} approximations to the correlation function. 
The zero-dimensional $\phi^4$ model is a probability distribution over a single real variable \( \phi \), with the probability density:
\begin{align}\label{0d_phi4_PDF}
\rho(\phi) \propto e^{-\frac{1}{2} m^2 \phi^2 - \frac{g}{4!}\phi^4}
\end{align}
where $m,g>0$. For this interacting field, we aim to compute the two-point correlation function of the field, which is defined by its second moment:
$G = \langle \phi^2 \rangle = \frac{1}{Z} \int_{-\infty}^{+\infty} \phi^2\, e^{-\frac{1}{2} m^2 \phi^2 - \frac{g}{4!} \phi^4} dq
$. The result admits an analytical solution:
\begin{align}\label{od_G_analytic}
G = \frac{4}{m^4} \rho \left[ \frac{K_{3/4}(\rho)}{K_{1/4}(\rho)} - 1 \right],
\end{align}
where $\rho = \frac{3m^4}{4g}$ and \( K_{1/4}(\cdot) \), \( K_{3/4}(\cdot) \) are modified Bessel functions of the second kind. If we fix $m=1$, the function $G=G(g)$ can be approximated using two perturbative expansions. On the one hand, as $g\rightarrow 0$, the weak coupling expansion for $G(g)$ can be readily obtained by expanding $e^{-\frac{g}{4!}\phi^4}$ and taking the ensemble average with respect to the Gaussian weight, i.e., following the standard many-body perturbation theory (MBPT) procedure \cite{parisi1988statistical,brown2015two}. On the other hand, the strong coupling expansion as $g\rightarrow +\infty$ can be derived in a similar manner by instead expanding the Gaussian factor \( e^{-\frac{1}{2} \phi^2} \) and averaging with respect to the non-Gaussian weight \( e^{-\frac{g}{4!} \phi^4} \). This leads to the following WCE and SCE series for $G(g)$:
\begin{align}
G(g)
=&1 - \frac{g}{2} + \frac{2g^2}{3} - \frac{11g^3}{8} + \frac{34g^4}{9} + \cdots 
\label{0d_wce}\\
\begin{split}
G(g) =& \left[\frac{2\sqrt{6}\,\Gamma\left(\frac{3}{4}\right)}{\Gamma\left(\frac{1}{4}\right)}\right]\frac{1}{\sqrt{g}}\\
+& 12\left[\left(\frac{\Gamma\left(\frac{3}{4}\right)}{\Gamma\left(\frac{1}{4}\right)}\right)^2 - \frac{\Gamma\left(\frac{5}{4}\right)}{\Gamma\left(\frac{1}{4}\right)}\right]\frac{1}{g} 
+\cdots
\end{split}\label{0d_sce}
\end{align}
The high-order expansion coefficients can be computed numerically using recursive formulas (see the Supplemental Material (SM)~\cite{suppl}). By setting $\tilde g =\sqrt{g}$ and reformulating \eqref{0d_wce} and \eqref{0d_sce} into series expansions with respect to $\tilde g$, we can employ the two-point Pad\'e approximant \cite{bender2013advanced} to construct global approximations to \( G(\tilde{g}) \). Specifically, the expansion ansatz is:
\begin{equation}\label{eqn:rational_ansatz}
\begin{aligned}
P_{[N/N+1]}(\tilde{g})
= \frac{A_0 + A_1 \tilde{g} + \cdots + A_N \tilde{g}^N}
       {1 + B_1 \tilde{g} + \cdots + B_{N+1} \tilde{g}^{N+1}},
\end{aligned}
\end{equation}
where the coefficients \( \{A_m, B_{m+1}\}_{m=0}^N \) are determined by matching the series expansions of the rational function ansatz \eqref{eqn:rational_ansatz} with WCE \eqref{0d_wce} and SCE \eqref{0d_sce} as:
\begin{equation}\label{eqn:matching}
\begin{aligned}
P_{[N/N+1]}(\tilde{g})
\sim
\begin{dcases}
a_0 + a_1 \tilde{g} + \cdots + a_N \tilde{g}^N, & \tilde{g} \to 0,\\[4pt]
b_1 \tilde{g}^{-1} + \cdots + b_{N+1} \tilde{g}^{-N-1}, & \tilde{g} \to +\infty.
\end{dcases}
\end{aligned}
\end{equation}
The numerical results are summarized in Fig.~\ref{fig:Gii_log}, which demonstrates that the two-point Pad\'e approximants converge to the exact function \( G(\tilde g) \) \emph{uniformly} (i.e., the error gets smaller as $N$ gets larger) and \emph{globally} (i.e., it holds for all $\tilde g$) for the cases studied. 

We note that other interpolation methods can also be used. For instance, one can directly interpolate on approximated function values obtained near the two endpoints using the truncated WCE/SCE series. Moreover, the two-point Pad\'e ansatz also admits various generalizations. 
In particular, one can freely choose the matching orders of WCE and SCE and construct a general \emph{Pad\'e-\(w_r\)-\(s_{2N+1-r}\)} scheme using the same rational function ansatz  ~\eqref{eqn:rational_ansatz}, where \(r\) and \(2N{+}1{-}r\) denote the matching orders of the WCE and SCE series, respectively.
The two expansion points can also be taken as \(\tilde{g} = 0\) and \(0 \ll \tilde{g}=\tilde g_0< +\infty\), leading to what we shall refer to as the \emph{Pad\'e–Taylor expansion} in the remainder of this work. 
Detailed derivations of these variants are provided in the SM~\cite{suppl}.

\paragraph{Strong-coupling expansion.}
\begin{figure}
\includegraphics[width=8.5cm]{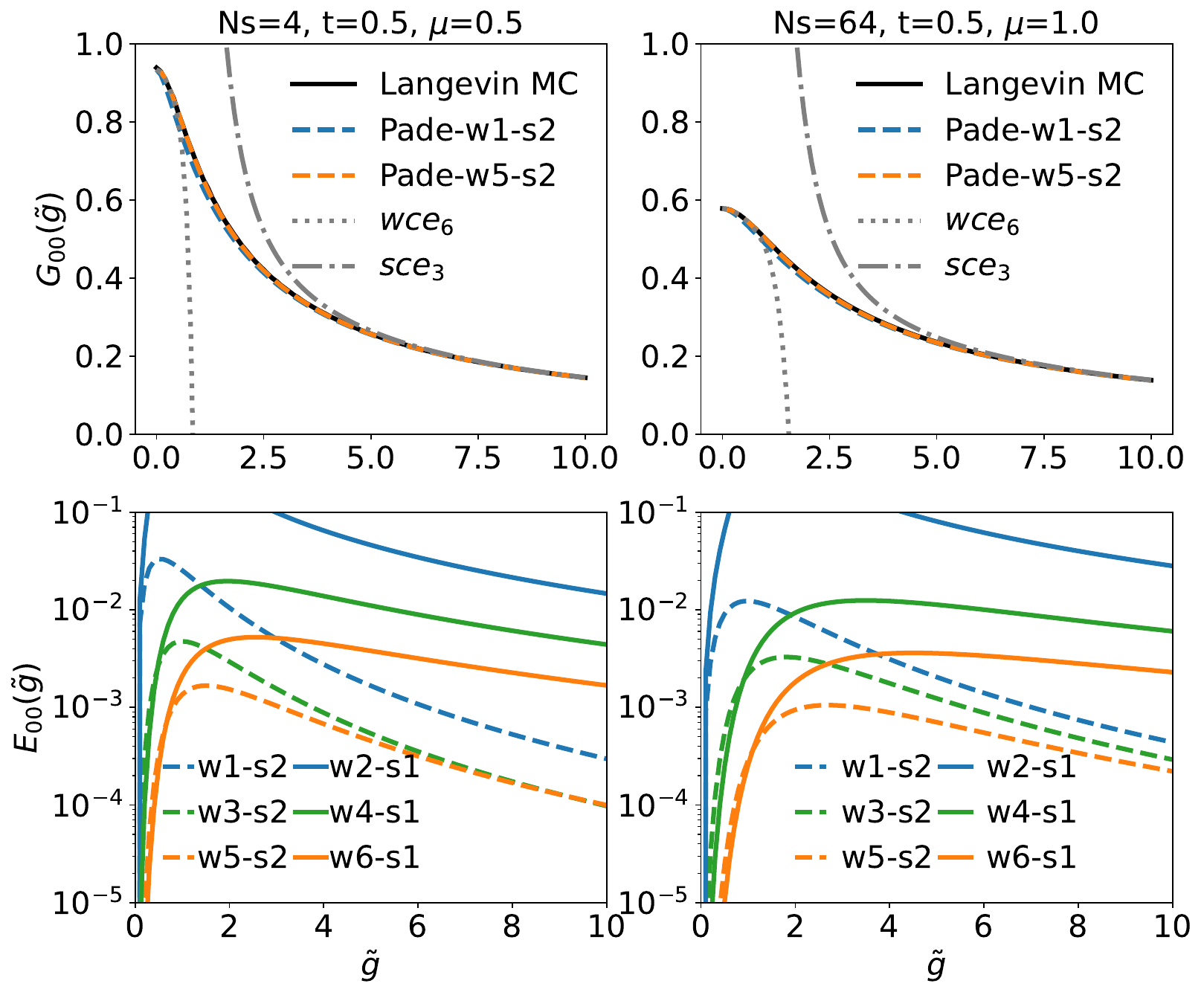}
\caption{\textbf{(First row)} Correlation function \( G_{00}(\tilde{g}) \) for the 1D lattice \( \phi^4 \) model with \( N_s = 4, 64\), computed using various Pad\'e-\( w_r \)-\( s_{2N+1-r} \) schemes and compared with Langevin Monte Carlo simulations (numerical accuracy $\sim 10^{-3}$). 
\textbf{(Second row)} Approximation error of the Pad\'e schemes. 
To better illustrate the numerical convergence, the error curves for different Pad\'e approximates are plotted relative to the highest-order result that we could obtain: Pad\'e-\( w_6 \)-\( s_3 \).
}
\label{fig:Gij_1d_phi4}
\end{figure}
The zero-dimensional $\phi^4$ example demonstrates the potential effectiveness of the two-point Pad\'e scheme in constructing global approximations of correlation functions for strongly interacting fields. Once the weak- and strong-coupling expansions are available, forming the corresponding two-point Pad\'e approximation is straightforward. In our construction, the WCE is always provided by standard MBPT \cite{parisi1988statistical,zinn2021quantum,negele2018quantum}. Thus, the main challenge reduces to developing the SCE for interacting fields, which constitutes the most technically demanding part of our approach.

For a general interacting field, the SCE is valid in the limit of coupling strength $g \to +\infty$ and can be, in principle, obtained by treating the interaction part of the Hamiltonian as unperturbed while expanding the non-interacting part in the path integral. However, for non-zero-dimensional fields, proceeding in this naive way leads to an expansion series that quickly becomes intractable as Wick’s theorem is no longer valid for a non-quadratic unperturbed Hamiltonian. To obtain a systematic treatment of SCE, we restrict our attention to fields containing only local quartic nonlinearities, exemplified by the lattice $\phi^4$ field and the Hubbard model discussed below. For these systems, the Hubbard--Stratonovich transformation together with the techniques introduced by Pairault \textit{et al.}~\cite{pairault1998strong, pairault2000strong} and ourselves enable us to develop a SCE framework that treats the $\phi^4$ field and the Hubbard model on an equal footing. This part of the derivation involves a series of combinatorial techniques which, due to their technical complexity, will be presented elsewhere.  In this letter, we only make use of the final SCE results, which are summarized in the SM~\cite{suppl}.

\paragraph{Applications.}\- 
Let us first consider the \( \phi^4 \) field theory model on a one-dimensional chain of length $N_s$, where the Hamiltonian is given by:
\begin{align}\label{lattice_phi_4}
\mathcal{H} = \sum_{ij} \frac{t_{ij}}{2}(\phi_i - \phi_j)^2 + \sum_i \frac{\mu}{2} \phi_i^2 + \frac{g}{4!} \phi_i^4.
\end{align}
In \eqref{lattice_phi_4}, \( t_{ij} \) denotes the nearest-neighbor interaction, subject to periodic boundary conditions. Here we only focus on the 1D case with \( \mu, g > 0 \) and the generalization of our approach to systems with arbitrary dimension and $\mu<0$ is  straightforward \footnote{At this stage, we do not consider the continuum limit of \eqref{lattice_phi_4}, for which renormalization is needed. We also do not investigate the critical behavior as \( \mu \to 0 \). These aspects require separate analysis and will be pursued as independent research directions in future work.}. The quantity of interest is 
the two-point correlation function $G_{ij}:=\langle\phi_i\phi_j\rangle
=\frac{
\int D[\phi]e^{-\H[\phi]}\phi_i\phi_j
}{\int D[\phi]e^{-\H[\phi]}}
$. 
Using the standard MBPT to obtain the WCE for $G_{ij}(\tilde g)$ with $\tilde g=\sqrt{g}$, together with the SCE over $1/\tilde{g}$ ~\cite{suppl}, we apply the two-point Pad\'e expansion to approximate each matrix element of $G_{ij}(\tilde g)$.
Fig.~\ref{fig:Gij_1d_phi4} shows the numerical results for the diagonal $ G_{ii}(\tilde{g})$, where the two-point Pad\'e approximation exhibits global convergence (top panels) and uniform convergence (bottom panels). This behavior is consistent for both small ($N_{s}=4$, left column) and large ($N_{s}=64$, right column) lattices.  Similar convergence behavior is also observed for the off-diagonal components, which are provided in the SM~\cite{suppl}.

\begin{figure}
\centering
\includegraphics[width=8.2cm]{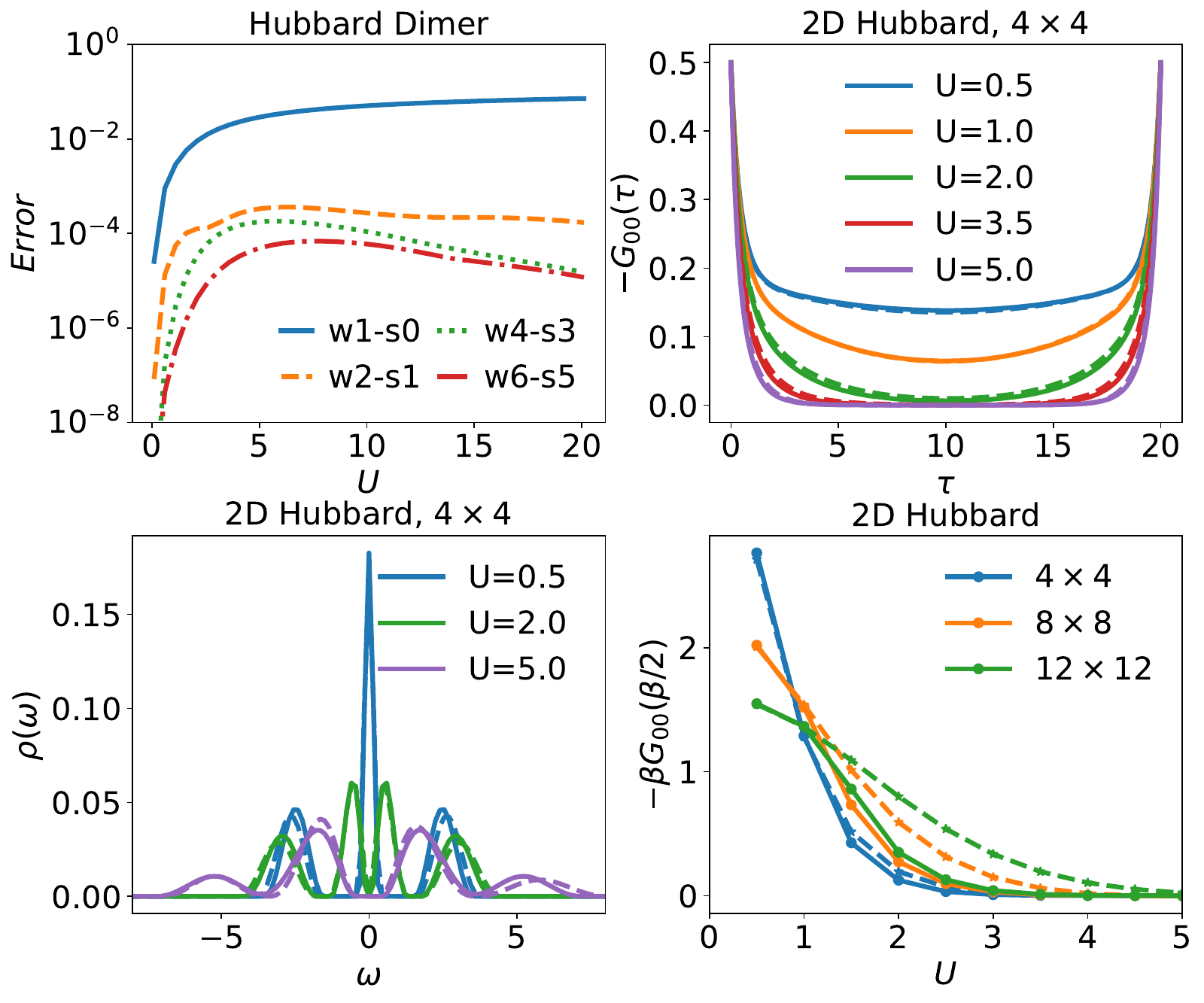}
\caption{\textbf{Top-left:} Pad\'e approximation error of the Matsubara Green's function
\(\frac{1}{N}\sum_{n=1}^{N}\bigl|G_{00}(i\omega_n)-G^{\mathrm{P}}_{00}(i\omega_n)\bigr|
\)
for the Hubbard dimer, using the analytic \(G_{00}(i\omega_n)\) as the reference. \textbf{Other panels:} Approximation results for 2D Hubbard model; solid lines show AFQMC reference and dashed lines show the Pad\'e-\(w_2\)-\(s_1\) approximation results. All simulations are at $t=1.0,\beta=20$. The Pad\'e-Taylor expansion interpolates between two points: $U=0$ and $U=10$. }
\label{fig:Gii_hubbard}
\end{figure}
%
%
%
For the next application, we consider the Hubbard model, where the Hamiltonian is given by:
\begin{align*}
    \H = -\sum_{ij\sigma}t_{ij}c_{i\sigma}^{\dagger}c_{j\sigma}+\sum_{i}Un_{i\uparrow}n_{i\downarrow}-\mu\sum_{i\sigma}n_{i\sigma}
\end{align*}
where \( t_{ij} \) denotes the nearest-neighbor hopping, and a periodic boundary condition is imposed. We restrict ourselves to the half-filling case where $\mu=U/2$ and aim to calculate the single-particle Green's function: $G_{ij,\sigma}(\tau) = -\langle \mathcal T c_{i\sigma}(\tau) c_{j\sigma}^\dagger(0) \rangle $. The Pad\'e interpolation is performed in frequency space, and the imaginary-time Green's function is obtained via the Fourier transform. Again, the WCE is obtained following the standard MBPT procedure~\cite{stefanucci2013nonequilibrium,fetter2012quantum}.
The SCE is obtained using procedures similar to those used for the $\phi^4$ theory~\cite{suppl,metzner1991linked}. However, unlike the $\phi^4$ case, the expansion here is carried out with respect to the hopping amplitude $|t_{ij}|$ rather than $1/U$, \textit{i.e.}, considering the limit $|t_{ij}| \to 0$ instead of $U \to \infty$ 
\footnote{These two limits proves to be different for calculating the imaginary time Green's function since the naive transformation $t\rightarrow t/U$ will accompanied with a change of inverse temperature $\beta\rightarrow \beta U$.
}.
This introduces additional difficulty when matching the WCE and SCE series through the Pad\'e approximation, since the rational function ansatz involves only a single variable $U$. To resolve this, a reformulation of the SCE series is introduced in the SM~\cite{suppl}, which ultimately enables the construction of the Pad\'e-Taylor expansion using $U$ as the only expansion variable for any fixed $\omega_n$ value. 

We first benchmark our numerical results on the Hubbard dimer. The Hubbard dimer Hamiltonian is easily diagonalizable, yielding an analytic solution $G(U, i\omega_n)$~\cite{suppl}. From this solution, we can directly extract the WCE/SCE series via differentiation rather than using diagrammatics and construct high-order two-point Pad\'e approximants \footnote{In this test, we temporarily set aside the diagrammatic computation problem of WCE/SCE and focus on validating the convergence property of the two-point Pad\'e expansion. The equivalence of the WCE/SCE series obtained by differentiation and diagrammatics for the Hubbard dimer will be detailed in SM ~\cite{suppl}}. 
As shown in the top-left panel of Fig.~\ref{fig:Gii_hubbard}, the approximants show clear numerical convergence with increasing order. For the two-dimensional Hubbard model on a square lattice, the WCE and SCE must be evaluated diagrammatically, which we have carried out up to second order. Getting higher-order expansion series is a separate research topic. One may for example employ diagrammatic QMC methods to obtain the
WCE~\cite{ferrero2023diagrammatic} and develop similar algorithms for the SCE. Nevertheless, we find that even at such low orders, the two-point Pad\'e expansion already provides a reasonable approximation to the single-particle Green’s function over a wide range of \( U/t \) values. 
The results are summarized in Fig.~\ref{fig:Gii_hubbard}, where a Pad\'e-Taylor interpolation for the Green's function at $U=0$ and $U=10$ is performed. 
Specifically, for the \( 4\times4 \) lattice, we computed the onsite Green’s function \( G_{ii}(\tau) \) and the density of states \( \rho(\omega) \) via analytic continuation, and compared the results with AFQMC benchmarks~\cite{qin2016benchmark}. 
The Pad\'e-\( w_2 \)-\( s_1 \) approximation captures the opening of the spectral gap as well as the spread of the bands (see the bottom-left panel). 
We further performed calculations for larger lattices and examined \( -\beta G_{ii}(\tau=\beta/2) \sim A(\omega=0) \), a commonly used indicator of the metal-to-insulator transition at low temperatures. The simulation result shows that the accuracy of the approximation decreases with increasing lattice sizes, indicating that higher expansion orders are required for quantitative agreement. It awaits further investigation how the convergence rate scales as the lattice size \( L \to \infty \).
Since the WCE is known to exhibit insensitivity to lattice size at sufficiently high expansion orders~\cite{ferrero2023diagrammatic}, one may expect a similarly favorable behavior for the SCE and hence the applicability of the two-point Pad\'e expansion.

\begin{figure}[t]
\centering
\includegraphics[width=7cm]{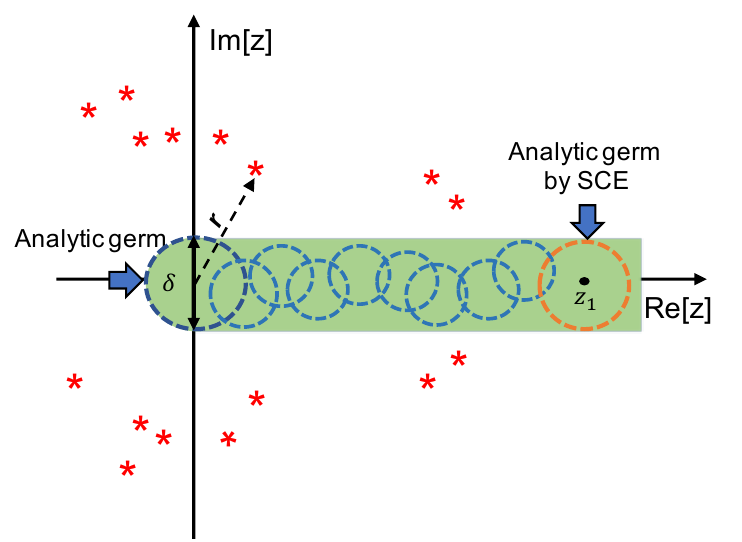}
\caption{Schematic illustration of the convergence mechanism via analytic continuation. 
If \( f(z) \) is pole-free within a \(\delta\)-tube enclosing \( \{0\} \cup \mathbb{R}^{+} \), it can be obtained by successive AC from the germ at \( z=0 \) to the whole tube, thereby recovering \( f(x) \) for \( x \ge 0 \). This extrapolation procedure may converge slowly for large \( x \) (e.g. using one-point Pad\'e expansion); however, knowing an additional germ at \( 0<z_1\leq +\infty \) allows interpolative AC between the two germs that may converge rapidly to the original function (e.g. using two-point Pad\'e expansion). 
Red asterisks (\textcolor{red}{*}) mark the singularities of \( f(z) \), and \( r \) denotes the Taylor-series convergence radius at \( z=0 \). Note that at two ends, the series convergence regimes may be much larger and the displayed analytic germs only show their overlapping with the \(\delta\)-tube.
}
\label{fig:AC_illustration}
\end{figure}

\paragraph{Why is the convergence possible?} 
Despite the complexity of obtaining the SCE for a general field-theoretic model, what makes this exploration worthwhile is the particularly striking feature that the two-point interpolation not only yields \emph{accurate} approximations to the Green’s function, but also exhibits clear numerical \emph{convergence} as the expansion order increases. This convergence property makes the approach especially appealing, as it offers a pathway to obtain quantitatively accurate results for strongly correlated systems in a systematically improvable and computationally scalable manner since all the calculations are based on bare perturbation results. 

Proving the convergence of Pad\'e approximants is notoriously challenging~\cite{baker1981pade,stahl1997convergence,buslaev2013convergence}, let alone establishing such results for correlation functions of general statistical or quantum fields. 
Nevertheless, we can provide heuristic arguments to explain why such global convergence is \emph{possible}. 
Mathematically, this amounts to asking given information about a function \( f(x) \) at one or two points, whether it is possible to reconstruct the entire function for \( x \in \mathbb{R}^+ \). 
For a real variable \( x \), this reconstruction is generally impossible due to non-uniqueness of solution. 
However, if \( f(x) \) admits an analytic continuation to the complex plane, the situation is entirely different. 
As illustrated in Fig.~\ref{fig:AC_illustration}, the analytically continued function \( f(z) \) may possess poles or branch points in \( \mathbb{C} \). But if all the singularities lie away from the real axis—say, outside a small \(\delta\)-tube containing \( \{0\} \cup \mathbb{R}^{+} \)—then \( f(z) \) is \emph{uniquely} defined and analytic within this domain. 
Consequently, starting from a locally convergent power series at \( z=0 \) (a \emph{germ}), one can recover this globally defined analytic function through successive analytic continuation (AC)\cite{ahlfors1979complex}.
Pad\'e expansion happens to be a practical numerical tool that can effectively perform such AC \cite{george2012quantitative,baker1981pade}. Moreover, incorporating information from two points, i.e., employing two-point Pad\'e expansion, effectively transforms the \emph{extrapolative} AC problem into an \emph{interpolative} AC problem, thereby yielding a natural global approximation to the correlation function that may converge rapidly. 

The numerical convergence of the two-point Pad\'e expansion observed for the lattice \( \phi^4 \) fields and the Hubbard model can be plausibly explained by this mechanism. In particular, for the Hubbard dimer, we can explicitly determine all poles of the analytically continued Green’s functions \( G(U,i\omega_n) \), with \( U\in\mathbb{C} \) ~\footnote{Here we particularly emphasize that the AC for the Matsubara Green's function $G(U,i\omega_n)$ of the Hubbard model is with respect to the interaction strength $U$ for each $i\omega_n$, {\em not} the AC $i\omega_n\rightarrow \omega+i0^+$ that are used to find the spectral function. }, and verify that they indeed exhibit the structures shown in Fig.~\ref{fig:AC_illustration} (the analysis for $\phi^4$ example is similar but more delicate since the original WCE series is only asymptotic. See SM~\cite{suppl} for details). Guided by the Lee--Yang theory for the Ising model~\cite{lee1952statistical,yang1952statistical}, we expect that the pole-free property of the correlation function near the real axis persists for finite systems at finite temperature or for infinite systems at high temperatures whenever the phase transition does not occur and the correlation functions remain analytic. Even in cases involving phase transitions, the two-point Pad\'e expansion can still be used to probe the system’s behavior near the critical point by constructing the expansion at finite \( L \) and then gradually approaching the thermodynamic limit \( L \rightarrow \infty \) until the finite-size effect becomes negligible. With that being said, how precisely the strong correlation regime is approached and resolved, and how well singularities at phase transitions are approximated, is an interesting topic for future study.

\paragraph{Conclusion and outlook.}
We developed a interpolation approach to construct global approximations to the Green’s function using the two-point Pad\'e expansion that bridges the weak and strong coupling expansions of interacting field theories. When applied to the lattice \( \phi^4 \) field theory and the Hubbard model, the two-point Pad\'e approximates exhibit clear convergence toward the exact correlation function across a broad range of interaction strengths. This convergence behavior can be understood through the analytic continuation of the Green’s function and its pole structure in the complex plane. As a new non-perturbative framework, this approach can be combined with CDet and other diagrammatic QMC techniques \cite{ferrero2023diagrammatic,Van_Houcke_2010,rossi2017determinant} to obtain higher-order approximations for various correlation functions and potentially a wider class of thermodynamic quantities. We believe that this work paves a promising pathway toward accurate and efficient simulations of strongly correlated systems, potentially offering new insights into long-standing problems in the relevant field.

\vspace{0.5cm}
We thank S. Zhang’s group for providing us the AFQMC results for the 2D Hubbard model.  Y. Zhu acknowledges valuable discussions with colleagues, including L. Lin, S. Zhang, Z. Huang, P. Xie, P. Rosenberg and V. Vlcek. Y. Yu acknowledges a helpful discussion with Jiuci Xu. 


This material is based upon work supported by the U.S. Department of Energy, Office of Science, Office of Advanced Scientific Computing Research and Office of Basic Energy Sciences, Scientific Discovery through Advanced Computing (SciDAC) program under Award Number DE-SC0022198. This work is also supported by the Center for Computational Study of Excited-State Phenomena in Energy Materials (C2SEPEM) at the Lawrence Berkeley National Laboratory, which is funded by the U.S. Department of Energy, Office of Science, Basic Energy Sciences, Materials Sciences and Engineering Division, under Contract No. DE-AC02-05CH11231, as part of the Computational Materials Sciences Program. This research used resources of the National Energy Research Scientific Computing Center, a DOE Office of Science User Facility supported by the Office of Science of the U.S. Department of Energy under Contract No. DE-AC02-05CH11231 using NERSC award ASCR-ERCAP m1027. Y.Yu was supported by the National Science Foundation under Grant No. NSF DMR 2401159. EK was supported by the U.S. Department of Energy (DOE) under Contract No. DE-AC02-05CH11231 through the Office of Advanced Scientific Computing Research Accelerated Research for Quantum Computing Program, FAR-QC. EG acknowledges funding from the European Research Council (ERC) under Advanced Grant No. 101142136 (Quantum Algorithms).

\bibliographystyle{apsrev4-2}
\bibliography{Derivation}
\end{document}


\title{Supplementary Materials\\ Global approximations to correlation functions of strongly interacting quantum field theories}

\maketitle

\section{Two-Point Pad\'e approximation}
\subsection{Standard two-point Pad\'e approximation}
Suppose a scalar function \( f(x) \) is well-defined on the semi-infinite domain \( x \in [0, +\infty) \), and that its asymptotic behavior is known at both endpoints, \( x = 0 \) and \( x = +\infty \). Specifically, we assume that \( f(x) \) admits the following series expansions:
\begin{align}
f(x) &= a_0 + a_1 x + a_2 x^2 + \cdots, \qquad x \to 0, 
\label{wce_1}\\[4pt]
f(x) &= \frac{b_1}{x} + \frac{b_2}{x^2} + \cdots, \qquad x \to +\infty,
\label{sce_1}
\end{align}
where we have taken \( b_0 = 0 \) for simplicity; the case \( b_0 \neq 0 \) will be discussed separately in Section~\ref{sec:2pade_b0_neq0}. To construct a global approximation that interpolates between these two asymptotic regimes, we employ an \( N \)-th order two-point Pad\'e approximant, denoted by \( P_{[N/N+1]}(x) \). The two-point Pad\'e approximation is represented as a rational function of the form
\begin{equation}
\begin{aligned}
P_{[N/N+1]}(x)
= \frac{R_N(x)}{S_{N+1}(x)}
= \frac{A_0 + A_1 x + \cdots + A_N x^N}{1 + B_1 x + \cdots + B_{N+1} x^{N+1}},
\end{aligned}
\end{equation}
where the coefficients \( \{A_n, B_{n+1}\}_{n=0}^N \) are determined by matching with the known series expansions in Eqs.~\eqref{wce_1}–\eqref{sce_1} up to the desired order. This requires the series expansion of \( P_{[N/N+1]}(x) \) to satisfy the following condition:
\begin{equation}\label{Pad\'e_matching_cond}
\begin{aligned}
R_N(x)-S_{N+1}(x)\sum_{n=0}^{\infty} a_nx^n &\sim O(x^{N+1}),\quad x\rightarrow 0, \\
R_N(x)-S_{N+1}(x)\sum_{n=1}^{\infty} b_nx^{-n} &\sim O(x^{-N-2}) ,\quad x\rightarrow+\infty.
\end{aligned}
\end{equation}
The above matching condition leads to the following linear systems for $\{A_n,B_{n+1}\}_{n=0}^N$:
\begin{align*}
\begin{dcases}
    A_n &=\sum_{j=0}^na_{n-j}B_j,\qquad 0\leq n\leq N, \\
    A_n &= \sum_{j=n}^NB_{j+1}b_{j-n+1},\qquad 0\leq n\leq N.
\end{dcases}
\end{align*}
where $B_0=1$. These two equations can be equivalently written as $\sum_{j=0}^na_{n-j}B_j=\sum_{j=n}^NB_{j+1}b_{j-n+1}$ and $A_n =\sum_{j=0}^na_{n-j}B_j$, or using the following matrix equation:
\begin{equation}
\begin{bmatrix}
b_1 & b_2 & \cdots & b_{N+1} \\
-a_0 & b_1 & \cdots & b_{N} \\
-a_1 & -a_0 & \cdots & b_{N-1} \\
\vdots & \vdots & \ddots & \vdots \\
-a_{N-1} & -a_{N-2} & \cdots & b_1
\end{bmatrix}
\begin{bmatrix}
B_1 \\
B_2 \\
\vdots \\
B_{N+1}
\end{bmatrix}
=
\begin{bmatrix}
a_0 \\
a_1 \\
\vdots \\
a_{N}
\end{bmatrix},
\end{equation}

\begin{equation}
\begin{bmatrix}
a_0 & 0 & \cdots & 0 \\
a_1 & a_0 & \cdots & 0 \\
\vdots & \vdots & \ddots & \vdots \\
a_{N} & a_{N-1} & \cdots & a_0
\end{bmatrix}
\begin{bmatrix}
1 \\
B_1 \\
\vdots \\
B_{N}
\end{bmatrix}
=
\begin{bmatrix}
A_0 \\
A_1 \\
\vdots \\
A_{N}
\end{bmatrix}.
\end{equation}

\subsection{Generalized two-point Pad\'e approximation}
Unlike the one-point Pad\'e approximation \( P_{[N/M]}(x) \), where the polynomial degrees in the numerator (\(N\)) and denominator (\(M\)) can be chosen freely, the two-point Pad\'e approximation requires a fixed rational form: \( P_{[N/N+1]}(x) \) (or \( P_{[N/N]}(x) \) when \( b_0 \neq 0 \)), to ensure the solvability of the linear systems \cite{roy2009global}. However, we can freely choose the matching orders of the series at $x\to0$ (weak-coupling expansion [WCE]) and  the series at $x\to\infty$ (strong-coupling expansion [SCE]). This leads to the  {\em generalized two-point Pad\'e approximation}, denoted by Pad\'e-$w_r$-$s_{2N+1-r}$, where we match WCE up to order $r$ and SCE to the order $2N+1-r$, with $N\leq r\leq 2N$. Correspondingly, we have the new matching condition:
\begin{equation}\label{general_Pad\'e_match_cond}
\begin{aligned}
P_{[N/N+1]}(x)
\sim
\begin{dcases}
a_0+a_1x+\cdots +a_{r}x^{r},\quad x\rightarrow 0,\\
b_1x^{-1}+\cdots +b_{2N+1-r}x^{-2N-1+r},\quad x\rightarrow+\infty,
\end{dcases}
\end{aligned}
\end{equation}
which leads to the following linear systems for $\{A_n,B_{n+1}\}_{n=0}^N$:
\begin{align*}
\begin{dcases}
    \sum_{i=0}^ja_{j-i}B_i = A_j, &\qquad 0\leq j\leq N,\\
    \sum_{i=0}^{N+1} a_{j-i} B_i = 0, &\qquad N+1\leq j \leq r,\\
   \sum_{i=0}^jB_{N-i+1}b_{j-i+1}= A_{N-j}, &\qquad 0\leq j\leq 2N-r.
\end{dcases}
\end{align*}
For any $N\leq r\leq 2N$,  it is equivalent to solving the corresponding matrix equations:
\begin{equation}\label{matrix_eqn_b0}
\begin{aligned}
\begin{bmatrix}
[\bm C_1^r]_{(2N+1-r)\times (N+1)}\\
[\bm C_2^r]_{(r-N)\times (N+1)}
\end{bmatrix}
\begin{bmatrix} B_1 \\ B_2 \\ \vdots \\ B_{N+1} \end{bmatrix} 
=
\begin{bmatrix}
a_{r-N} \\
a_{r-N+1} \\
\vdots \\
a_{r}
\end{bmatrix}
,\quad 
\begin{bmatrix}
a_0 & 0 & \cdots & 0 \\
a_1 & a_0 & \cdots & 0 \\
\vdots & \vdots & \ddots & \vdots \\
a_{N} & a_{N-1} & \cdots & a_0
\end{bmatrix}
\begin{bmatrix}
1 \\
B_1 \\
\vdots \\
B_{N}
\end{bmatrix}
=
\begin{bmatrix}
A_0 \\
A_1 \\
\vdots \\
A_{N}
\end{bmatrix}
\end{aligned},
\end{equation}
where $[\bm C_1^r]_{(2N+1-r)\times (N+1)}$ is the last $2N+1-r$ row of matrix $\bm C_1$, and $[\bm C_2^r]_{(r-N)\times (N+1)}$ is the first $r-N$ row of matrix $\bm C_2$. $\bm C_1$ and $\bm C_2$ are defined as:
\begin{align*}
\bm C_1 =
\begin{bmatrix}
b_1 & b_2 & \cdots & b_{N+1} \\
-a_0 & b_1 & \cdots & b_{N} \\
-a_1 & -a_0 & \cdots & b_{N-1} \\
\vdots & \vdots & \ddots & \vdots \\
-a_{N-1} & -a_{N-2} & \cdots & b_1
\end{bmatrix}
,\qquad
\bm C_2
=
\begin{bmatrix}
-a_N & -a_{N-1} & \cdots & -a_0 \\
-a_{N+1} & -a_N & \cdots & -a_{1} \\
-a_{N+2} & -a_{N+1} & \cdots & -a_2 \\
\vdots & \vdots & \ddots & \vdots \\
-a_{2N} & -a_{2N-1} & \cdots & -a_N
\end{bmatrix}.
\end{align*}
\subsection{$P_{[N/N]}(x)$ ansatz for $b_0\neq 0$}
\label{sec:2pade_b0_neq0}
All the strong-coupling expansions considered in this work have the leading-order term \( b_0 = 0 \); therefore, the two-point Pad\'e approximation follows the standard construction introduced in the previous section. 
For completeness, however, we also present the formulation of the two-point Pad\'e expansion for cases where \( b_0 \neq 0 \). When \( b_0 \neq 0 \), the appropriate Pad\'e ansatz must take the form \( P_{[N/N]}(x) \)~\cite{roy2009global}. 
The corresponding matching condition for the generalized Pad\'e approximation, denoted as Pad\'e-\( w_r \)-\( s_{2N-r-1} \), is given by
\begin{equation}\label{general_pade_match_cond_b0}
\begin{aligned}
P_{[N/N]}(x)
= \frac{A_0 + A_1 x + \cdots + A_N x^N}{1 + B_1 x + \cdots + B_N x^N} 
\sim
\begin{dcases}
a_0 + a_1 x + \cdots + a_r x^{r}, \qquad &x \to 0, \\[4pt]
b_0 + b_1 x^{-1} + \cdots + b_{2N-1-r} x^{-2N+1+r}, \qquad &x \to +\infty,
\end{dcases}
\end{aligned}
\end{equation}
where \( N \le r \le 2N \). This leads to the following set of equations for the Pad\'e coefficients $\{A_0,\cdots A_N; B_1,\cdots, B_{N}\}$:
\begin{align*}
    \begin{dcases}
\sum_{i=0}^{j} a_{j-i} B_i = A_j, &\quad 0\leq j\leq N,\\
\sum_{i=0}^{N} a_{j-i} B_i = 0, &\quad N+1\leq j \leq r,\\
\sum_{i=0}^{j} b_{j-i} B_{N-i} = A_{N-j}, &\quad 
0\leq j\leq 2N-r-1.
\end{dcases}
\end{align*}
The corresponding matrix equations are:
\begin{equation}\label{matrix_eqn_b0_neq}
\begin{bmatrix}
[\bm C_1^r]_{(2N-r)\times (N)}\\
[\bm C_2^r]_{(r-N)\times (N)}
\end{bmatrix}
\begin{bmatrix} B_1 \\ B_2 \\ \vdots \\ B_{N} \end{bmatrix} 
=
\begin{bmatrix}
a_{r-N+1} \\
a_{r-N+2} \\
\vdots \\
a_{r}
\end{bmatrix}
,\quad
\begin{bmatrix}
a_0 & 0 & \cdots & 0 \\
a_1 & a_0 & \cdots & 0 \\
\vdots & \vdots & \ddots & \vdots \\
a_{N} & a_{N-1} & \cdots & a_0
\end{bmatrix}
\begin{bmatrix}
1 \\
B_1 \\
\vdots \\
B_{N}
\end{bmatrix}
=
\begin{bmatrix}
A_0 \\
A_1 \\
\vdots \\
A_{N}
\end{bmatrix},
\end{equation}
where $[\bm C_1^r]_{(2N-r)\times (N)}$ is the last $2N-r$ row of matrix $\bm C_1$, and $[\bm C_2^r]_{(r-N)\times (N)}$ is the first $r-N$ row of matrix $\bm C_2$. $\bm C_1$ and $\bm C_2$ are defined as:
\begin{align*}
\bm C_1 =
\begin{bmatrix}
b_0-a_0 & b_1 & \cdots & b_{N-1} \\
-a_1 & b_0-a_0 & \cdots & b_{N} \\
-a_2 & -a_1 & \cdots & b_{N-1} \\
\vdots & \vdots & \ddots & \vdots \\
-a_{N-1} & -a_{N-2} & \cdots & b_0-a_0
\end{bmatrix}
,\qquad
\bm C_2
=
\begin{bmatrix}
-a_N & -a_{N-1} & \cdots & -a_1 \\
-a_{N+1} & -a_N & \cdots & -a_{2} \\
-a_{N+2} & -a_{N+1} & \cdots & -a_3 \\
\vdots & \vdots & \ddots & \vdots \\
-a_{2N} & -a_{2N-1} & \cdots & -a_N
\end{bmatrix}.
\end{align*}
For $r=N$, the corresponding matrix equations are:
\begin{equation}
\begin{bmatrix}
b_0-a_0 & b_1 & \cdots & b_{N-1} \\
-a_1 & b_0-a_0 & \cdots & b_{N} \\
-a_2 & -a_1 & \cdots & b_{N-1} \\
\vdots & \vdots & \ddots & \vdots \\
-a_{N-1} & -a_{N-2} & \cdots & b_0-a_0
\end{bmatrix}
\begin{bmatrix} B_1 \\ B_2 \\ \vdots \\ B_{N} \end{bmatrix} 
=
\begin{bmatrix}
a_{1} \\
a_{2} \\
\vdots \\
a_{N}
\end{bmatrix}
,\quad
\begin{bmatrix}
a_0 & 0 & \cdots & 0 \\
a_1 & a_0 & \cdots & 0 \\
\vdots & \vdots & \ddots & \vdots \\
a_{N} & a_{N-1} & \cdots & a_0
\end{bmatrix}
\begin{bmatrix}
1 \\
B_1 \\
\vdots \\
B_{N}
\end{bmatrix}
=
\begin{bmatrix}
A_0 \\
A_1 \\
\vdots \\
A_{N}
\end{bmatrix}.
\end{equation}
\subsection{Pad\'e-Taylor expansion}
In the main text, all the Hubbard model calculations used a weak coupling expansion at $x=0$ and a strong coupling expansion at $x=x_{0}$ ($0 < x_0 < +\infty $)
\begin{align}
f(x) &=\sum_{n=0}^{\infty} a_n x^n, \quad x \to 0, \label{Pade-taylor_z0} \\
f(x) &= \sum_{n=0}^{\infty} b_n (x - x_0)^n, \quad x \to x_0. \label{Pade-taylor_zx0}
\end{align}
These expansions allow for more flexibility in the form of the rational ansatz for Pad\'e approximation, which may now be chosen as a general Padé approximant of type \( P_{[N/M]} \):
\begin{align}\label{Pade-taylor_ansatz}
f(x) \approx P_{[N/M]}(x) = \frac{A_0 + A_1 x + \cdots + A_N x^N}{1 + B_1 x + \cdots + B_M x^M}.
\end{align}

To distinguish this method from the two-point Padé approximation introduced earlier, we refer to it as the \emph{Padé-Taylor expansion}.
The \( N+M+1 \) coefficients in \eqref{Pade-taylor_ansatz} are determined by matching terms from both series expansions \eqref{Pade-taylor_z0} and \eqref{Pade-taylor_zx0}. If the Padé-Taylor approximant matches the first \( r \) terms of the weak-coupling series, then the remaining \( N+M-r+1 \) degrees of freedom are fixed by matching the Taylor expansion around \( x_0 \). The matching condition now becomes:
\begin{equation}\label{PT_match_cond}
\begin{aligned}
P_{[N/M]}(x)
\sim
\begin{dcases}
a_0 + a_1 x + \cdots + a_r x^r, &\quad x \to 0 \\
b_0 + b_1 (x - x_0) + \cdots + b_{N+M - r - 1}(x - x_0)^{N+M - r - 1}, &\quad x \to x_0.
\end{dcases}
\end{aligned}
\end{equation}
Without loss of generality, here we assume $a_0,b_0\neq 0$. A system of linear equations can then be constructed to determine the Padé-Taylor coefficients \( \{A_0, \dots, A_N; B_1, \dots, B_M\} \) from the expansion coefficients \( \{a_0, \dots, a_r;\, b_0, \dots, b_{N+M-r-1}\} \). Due to flexibility in the choice of $N,M,r$, here we only consider the case $N\leq r\leq N+M$ and a generalization is straightforward. By direct computation, it can be shown that matching condition \eqref{PT_match_cond} leads to:
\begin{equation}\label{PT_matrix_match_don}
\begin{bmatrix}
a_0 & 0 & \cdots & 0 \\
a_1 & a_0 & \cdots & 0 \\
\vdots & \vdots & \ddots & \vdots \\
a_{r} & a_{r-1} & \cdots & a_0\\
\end{bmatrix}
\begin{bmatrix}
\bm B_M\\
0\\
\vdots\\
0
\end{bmatrix}
=
\begin{bmatrix}
\bm A_N\\
0\\
\vdots\\
0
\end{bmatrix}
,\quad 
\begin{bmatrix}
b_0 & 0 & \cdots & 0 \\
b_1 & b_0 & \cdots & 0 \\
\vdots & \vdots & \ddots & \vdots \\
b_{q} & b_{q-1} & \cdots & b_0\\
\end{bmatrix}
\begin{bmatrix}
D_0\\
D_1\\
\vdots\\
D_{q+1}
\end{bmatrix}
=
\begin{bmatrix}
C_0\\
C_1\\
\vdots\\
C_{q}
\end{bmatrix},
\end{equation}
where $\bm B_M=[1,B_1,\cdots, B_M]^T,\bm A_N=[A_0,\cdots, A_N]^T$, and $q=N+M-r-1$. Note that the matrix in the second equation is {not} a square matrix. In \eqref{PT_matrix_match_don}, the $C, D$ elements are defined by:
\begin{equation}\label{PT_matrix_match_don1}
\begin{bmatrix}
\binom{0}{0} & \binom{1}{0}x_0 & \cdots & \binom{N}{0}x_0^N\\
0 & \binom{1}{1}  & \cdots & \binom{N}{1}x_0^{N-1}  \\
\vdots & \vdots & \ddots & \vdots \\
0 & \cdots &\cdots & \binom{N}{N} \\
\end{bmatrix}
\begin{bmatrix}
A_0\\
A_1\\
\vdots\\
A_N
\end{bmatrix}
=
\begin{bmatrix}
C_0\\
C_1\\
\vdots\\
C_N
\end{bmatrix}
,\quad 
\begin{bmatrix}
\binom{0}{0} & \binom{1}{0}x_0 & \cdots & \binom{M}{0}x_0^M\\
0 & \binom{1}{1}  & \cdots & \binom{M}{1}x_0^{M-1}  \\
\vdots & \vdots & \ddots & \vdots \\
0 & \cdots &\cdots & \binom{M}{M} \\
\end{bmatrix}
\begin{bmatrix}
1\\
B_1\\
\vdots\\
B_{M}
\end{bmatrix}
=
\begin{bmatrix}
D_0\\
D_1\\
\vdots\\
D_{M}
\end{bmatrix}.
\end{equation}
Are the calculations for the Hubbard model in the main text uss the diagonal Pad\'e form where $N=M$. As an example, for $N=M=2$ and $r=2$, the resulting Pad\'e-Taylor ansatz Pade-$w_2$-$s_1$ matches the WCE up to the 2nd order and SCE up to the 1st order. Matching conditions \eqref{PT_matrix_match_don}-\eqref{PT_matrix_match_don1} can be more explicitly rewritten as the following matrix equations for $\bm A_N, \bm B_M$:
\begin{equation}
\begin{aligned}
A_0 = a_0,
\quad
&
\begin{bmatrix}
(a_0-b_0)x_0+a_1x_0^2 & (a_0-b_0)x_0^2 \\
(a_0-b_0)+(2a_1-b_1)x_0 & 2(a_0-b_0)x_0-b_1x_0^2 
\end{bmatrix}
\begin{bmatrix}
B_1 \\
B_2
\end{bmatrix}
=
\begin{bmatrix}
b_0-a_0-a_1x_0-a_2x_0^2 \\
b_1-a_1-2a_2x_0
\end{bmatrix}
\end{aligned},
\end{equation}
and 
\begin{equation}
\begin{aligned}
\begin{bmatrix}
A_1 \\
A_2
\end{bmatrix}
&=
\begin{bmatrix}
a_0 & 0 \\
a_1 & a_0 
\end{bmatrix}
\begin{bmatrix}
B_1 \\
B_2
\end{bmatrix}
+
\begin{bmatrix}
a_1 \\
a_2
\end{bmatrix}.
\end{aligned}
\end{equation}

\paragraph{Remark}
In this work, all two-point Padé expansion coefficients are obtained by solving linear systems for 
$A$s and $B$s.
However, this procedure may become numerically unstable especially at higher orders due to the emergence of spurious pole–zero pairs in the Padé approximants~\cite{trefethen2019approximation}. 
Indeed, we occasionally observed such instabilities at certain expansion orders in our examples. 
For these cases, we simply adjusted the matching order in the Padé ansatz—for example, switching from Padé-\(w_2\)-\(s_1\) to Padé-\(w_1\)-\(s_2\)—or tuned the value of \(x_0\) in the Padé–Taylor expansion to restore smooth interpolation. 
More advanced stabilization techniques, such as SVD-based Padé algorithms~\cite{gonnet2013robust}, can be incorporated to mitigate these issues.  
We leave such improvements for future work.

\paragraph{Remark}
All of the two-point Padé interpolation schemes discussed above can be extended to approximate matrix-valued functions, where the expansion coefficients become coefficient matrices. In many applications, such matrix-valued functions correspond to the Green’s (correlation) functions of lattice models and the matrix indices correspond to site indices. The two-point Padé expansion can be applied elementwise to each matrix entry or directly at each momentum point \(k\). For instance, the two-point Padé approximation for 2D Hubbard model considered in this work are approximated as:
\begin{equation}
\begin{aligned}
G(U,\boldsymbol{k},i\omega_n)
&\approx 
P_{[N/N]}(U,\boldsymbol{k},i\omega_n)
= \frac{R_N(U,\boldsymbol{k},i\omega_n)}{S_N(U,\boldsymbol{k},i\omega_n)} \\
&= \frac{A_0(\boldsymbol{k},i\omega_n)
+ UA_1(\boldsymbol{k},i\omega_n)
+ \cdots
+ U^N A_N(\boldsymbol{k},i\omega_n)}
{1
+ UB_1(\boldsymbol{k},i\omega_n)
+ \cdots
+ U^N B_N(\boldsymbol{k},i\omega_n)},
\end{aligned}
\label{U_infty_series}
\end{equation}
where the Padé coefficients 
\(\{A_i(\boldsymbol{k},i\omega_n)\}_{i=0}^{N}\) 
and 
\(\{B_j(\boldsymbol{k},i\omega_n)\}_{j=1}^{N}\) are
determined by the linear system associated with the Pad\'e-Taylor expansion for each $\bm k$ and $i\omega_n$.
\section{WCE and SCE for lattice $\phi^4$ field theory and Hubbard model}
\subsection{Zero-dimensional $\phi^4$ field}
We derive the recursive formula for calculating the WCE and SCE expansion coefficients for the zero-dimensional $\phi^4$ field. For WCE around $g=0$, we have 
\begin{align}
G(g) = \frac{1}{Z}\int_{-\infty}^{\infty} dqq^2 e^{-\frac{1}{2} q^2 - \frac{g}{4!} q^4}
\sim
\frac{\displaystyle\sum_{n=0}^{\infty}\frac{\left(-\frac{g}{4!}\right)^n}{n!}\langle q^{4n+2}\rangle}
{\displaystyle\sum_{n=0}^{\infty}\frac{\left(-\frac{g}{4!}\right)^n}{n!}\langle q^{4n}\rangle}
=\frac{\displaystyle\sum_{n=0}^{\infty}g^na_n}
{\displaystyle\sum_{n=0}^{\infty}g^nb_n}
=\sum_{n=0}^{\infty}g^nG^w_n.
\label{0d_wce}
\end{align}
According to the Cauchy product formula:
\begin{align*}
\left(\sum_{i=0}^{\infty}b_i\right)\left(\sum_{j=0}^{\infty}c_j\right) = \sum_{k=0}^{\infty}a_k,
\qquad \text{where}
\quad a_k=\sum_{l=0}^kb_lc_{k-l},
\end{align*}
the recursive formula for calculating $G_n^w$ is:
\begin{align}
G_n^w=
\begin{cases}
\frac{a_0}{b_0},\quad n=0,\\
\displaystyle\frac{1}{b_0}\left[a_n-\sum_{k=0}^n G_k^wb_{n-k}\right],\quad n>0.
\end{cases}
\end{align}
For SCE around $g=+\infty$, we have 
\begin{align}
G(g) = \frac{1}{Z}\int_{-\infty}^{\infty} dqq^2 e^{-\frac{m^2}{2} q^2 - \frac{g}{4!} q^4}
\sim
\frac{\displaystyle\sum_{n=0}^{\infty}\frac{\left(-\frac{m^2}{2}\right)^n}{n!}\langle q^{2n+2}\rangle^S}
{\displaystyle\sum_{n=0}^{\infty}\frac{\left(-\frac{m^2}{2}\right)^n}{n!}\langle q^{2n}\rangle^S}
=\frac{\displaystyle\sum_{n=0}^{\infty}g^{-\frac{2n+3}{4}}a_n}
{\displaystyle\sum_{n=0}^{\infty}g^{-\frac{2n+1}{4}}b_n}
=\sum_{n=0}^{\infty}g^{-\frac{n}{2}}G^s_n,
\label{0d_sce}
\end{align}
where $\langle\cdot\rangle^S$ represent the ensemble average with respect to the probability density $e^{-\frac{g}{4!}q^4}$. The recursive formula for calculating $G_n^s$ is:
\begin{align}
G_n^s=
\begin{cases}
\frac{a_0}{b_0},\quad n=0,\\
\displaystyle\frac{1}{b_0}\left[ a_n-\sum_{k=0}^n G_k^sb_{n-k}\right],\quad n>0.
\end{cases}
\end{align}
\subsection{Lattice $\phi^4$ field}
\label{app:SCE_1d_phi4}
The WCE for the lattice \( \phi^4 \) field can be obtained using standard MBPT~\cite{parisi1988statistical}, and the corresponding diagrammatic expansion up to third order is provided in \cite{brown2015two}. 
Getting the expansion series from Feynman diagrams is straightforward, we therefore omit the details here. 
On the other hand, the SCE series for the correlation function of the lattice $\phi^4$ model is given by:
\begin{align}
\mathbf{G} = \tilde{g}^{-1} \mathbf{G}_s^{(1)} + \tilde{g}^{-2} \mathbf{G}_s^{(2)} + \tilde{g}^{-3} \mathbf{G}_s^{(3)} + \cdots
\end{align}
where $\tilde g=\sqrt{g}$, and the leading-order term is given by
\begin{align}\label{sce_1d_1}
[\mathbf{G}_s^{(1)}]_{ij} = \frac{2W(1, \mathbf{0})}{W(\mathbf{0})} \, \delta_{ij},
\end{align}
with \( \delta_{ij} \) the Kronecker delta (i.e., the identity matrix), and \( W(j_1, \dots, j_N) \) is a fully  symmetric (scalar) function defined as:
\begin{align}
W(j_1, \dots, j_N) =
\frac{(4!)^{-\frac{2m+1}{4}}}{\prod_{n=1}^N j_{n}!}\prod_{n=1}^N
\Gamma\left(\frac{2j_n+1}{4}\right),
\end{align}
where \( m = j_1 + \cdots + j_N \) and \( N \) is the total number of lattice sites. The notation \( \mathbf{0} \) in the argument of \( W \) indicates that all remaining arguments are zero. This is well-defined due to the full symmetry of \( W \): permuting the arguments \( \{j_1, \dots, j_N\} \) does not affect its value. The next two orders in the expansion are given by:
\begin{equation}\label{sce_1d_23}
\begin{aligned}
[\mathbf{G}^{(2)}_{s}]_{ij}
&=\left(-12\frac{W(2,\bm 0)}{W(\bm 0)}
+6\frac{W(1,1,\bm 0)}{W(\bm 0)}\right)K_{ii}\delta_{ij}
-4\frac{W(1,1,\bm 0)}{W(\bm 0)}K_{ij}\\
[\mathbf{G}^{(3)}_{s}]_{ij}
&=\left(90\frac{W(3,\bm 0)}{W(\bm 0)}
-90\frac{W(2,1,\bm 0)}{W(\bm 0)}+30\frac{W(1,1,1,\bm 0)}{W(\bm 0)}\right)K_{ii}\delta_{ij}\\
&+
\left(
24\frac{W(2,1,\bm 0)}{W(\bm 0)}-12\frac{W(1,1,1,\bm 0)}{W(\bm 0)}
\right)[K_{ii}K_{ij}+K_{ij}K_{jj}+\delta_{ij}\sum_kK_{ik}K_{ki}]\\
&+8\frac{W(1,1,1,\bm 0)}{W(\bm 0)}\sum_kK_{ik}K_{kj}
\end{aligned}
\end{equation}
where \( K \) denotes the matrix associated with the Gaussian measure, defined via:
\begin{align*}
\langle \phi | K | \phi \rangle := \sum_{ij} \frac{t_{ij}}{2} (\phi_i - \phi_j)^2 + \sum_i \frac{\mu}{2} \phi_i^2.
\end{align*}

To get the numerical results presented in the main text, we computed the WCE and SCE series up to third order. 
Since the WCE is originally expanded in powers of \(g\), rewriting it in terms of \(\tilde{g} = \sqrt{g}\) yields a series up to sixth order, where all odd-order terms vanish. 
Consequently, the highest-order two-point Padé approximant used in our study is the Padé-\(w_6\)-\(s_3\) scheme.

\subsection{Hubbard dimer}
The Hubbard dimer Hamiltonian is given by:
\begin{align}
H = -t\sum_{\sigma}\left(c_{1\sigma}^{\dagger}c_{2\sigma}
+ c_{2\sigma}^{\dagger}c_{1\sigma}\right)
+ U\left(n_{1\uparrow}n_{1\downarrow}
+ n_{2\uparrow}n_{2\downarrow}\right)
- \mu\sum_{\sigma}(n_{1\sigma}+n_{2\sigma}),
\end{align}
where the chemical potential is set to $\mu=U/2$ to enforce half-filling. 
The Hamiltonian can be diagonalized exactly, yielding all eigenvalues and eigenstates of the system~\cite{stefanucci2013nonequilibrium}. Using these, the exact Matsubara Green’s function \(G_{ij}(i\omega_n)\) can be computed via the Lehmann representation. At low temperatures, only transitions involving the ground state in the half-filled (two-electron) sector \(H^{(2)}\) and states in the adjacent one- and three-electron sectors \(H^{(1)}\) and \(H^{(3)}\) contribute significantly. Thus, \(G_{ij}(i\omega_n)\) can be computed as
\begin{align}
G_{ij,\sigma}(i\omega_n) &=
\frac{1}{Z} \sum_{m,n}
\left( e^{-\beta E_m} + e^{-\beta E_n} \right)
\frac{
\langle m | c_{i\sigma} | n \rangle 
\langle n | c_{j\sigma}^{\dagger} | m \rangle
}
{ i\omega_n + E_m - E_n } \nonumber\\
&\approx
\sum_{m \in H^{(1)}}
\frac{
\langle m | c_{i\sigma} | 0 \rangle 
\langle 0 | c_{j\sigma}^{\dagger} | m \rangle
}
{ i\omega_n + E_0 - E_m }
+ 
\sum_{m \in H^{(3)}}
\frac{
\langle m | c_{j\sigma}^{\dagger} | 0 \rangle 
\langle 0 | c_{i\sigma} | m \rangle
}
{ i\omega_n + E_0 - E_m },
\label{G_lehmann}
\end{align}
where \(|0\rangle\) denotes the ground state in \(H^{(2)}\). In the limit of $\beta\rightarrow+\infty$, the analytic expression is simplified as:
\begin{equation}
\begin{aligned}\label{G_analytic_t0}
G_{ij,\sigma}(i\omega_n)=G_{ij,\bar\sigma}(i\omega_n) 
&= \frac{1}{8a^2} \bigg( (-1)^{ij}\frac{\big[1+\frac{4t}{(c - U)}\big]^2}{i\omega_n+t-\frac{c}{2}} 
+ \frac{\big[1-\frac{4t}{(c - U)}\big]^2}{i\omega_n-t-\frac{c}{2}} \bigg)\\
&+ \frac{1}{8a^2} \bigg( \frac{\big[1+\frac{4t}{(c - U)}\big]^2}{i\omega_n-t+\frac{c}{2}} 
+ \frac{(-1)^{ij}\big[1-\frac{4t}{(c - U)}\big]^2}{i\omega_n+t+\frac{c}{2}} \bigg) \\
&+\frac{3}{16}\left(\frac{1}{i\omega_n+t-\frac{U}{2}}+\frac{(-1)^{ij}}{i\omega_n-t-\frac{U}{2}}
+\frac{(-1)^{ij}}{i\omega_n-t+\frac{U}{2}}
+\frac{1}{i\omega_n+t+\frac{U}{2}}\right)
\end{aligned}
\end{equation}
where $a=\frac{\sqrt{32t^2+2(c-U)^2}}{c-U}$, $c=\sqrt{16t^2+U^2}$. 
\paragraph{WCE} The weak-coupling expansion of $G_{ij}(i\omega_n)$ can be obtained by taking the Talyor series expansion of \eqref{G_analytic_t0} at $U=0$. For any $i\omega_n$ and fixed $t$, this leads to:
\begin{align}\label{wce_taylor}
G_{ij}(i\omega_n) = \sum_{m}U^mG_{ij}^{(m)}(t,i\omega_n).
\end{align}
It can be shown that all odd terms vanish. When $m=0$, we have:
\begin{align}\label{HF_solution}
G^{(0)}_{ij}(i\omega_n) =\frac{1}{2}\left[(-1)^{ij}\frac{1}{b_1-2t}+\frac{1}{b_2+2t}\right],
\end{align}
where $b_1= i\omega_n+t$, $b_2=i\omega_n-t$. This happens to be the Hartree-Fock (HF) solution. For $m=2$, we have: 
\begin{equation}\label{wce_G24}
\begin{aligned}
G^{(2)}_{ij}(i\omega_n) &=
\frac{1}{2}\left[
(-1)^{ij}\frac{6t-b_1}{64t^2(b_1-2t)^2}
-\frac{6t+b_2}{64t^2(b_2+2t)^2}
+\frac{1}{64t^2(b_2-2t)}
+(-1)^{ij}\frac{1}{64t^2(b_1+2t)}
\right]
\end{aligned}
\end{equation}

\paragraph{SCE} The strong-coupling expansion of $G_{ij}(i\omega_n)$ can be obtained by taking the Taylor series expansion of \eqref{G_analytic_t0} at $t=0$. For any $i\omega_n$ and fixed $U$, this leads to:
\begin{align}\label{sce_taylor}
G_{ij}(i\omega_n) = \sum_{m}t^m\hat G_{ij}^{(m)}(U,i\omega_n).
\end{align}
For $m=0$, the coefficient is
\[
\hat G^{(0)}_{ij}(U,i\omega_n) =\frac{i\omega_n}{(i\omega_n)^2-\frac{U^2}{4}}\delta_{ij}.
\]
This happens to the atomic Green's function, i.e., the Green's function for Hubbard dimer when $t=0$. When $m=1,2$, we have 
\begin{equation}\label{sce_G12}
\begin{aligned}
\hat G^{(1)}_{ij}(U,i\omega_n) &=\frac{-(i\omega_n)^2}{[(i\omega_n)^2-\frac{U^2}{4}]^2}(1-\delta_{ij}),\\
\hat G^{(2)}_{ij}(U,i\omega_n) 
&=\left(\frac{(i\omega_n)^3}{[(i\omega_n)^2-\frac{U^2}{4}]^3}+\frac{3}{4}\frac{i\omega_nU^2}{[(i\omega_n)^2-\frac{U^2}{4}]^3}\right)\delta_{ij}.
\end{aligned}
\end{equation}
Here we only list the WCE/SCE series up to the second order. Higher-order terms can be obtained by evaluating the Taylor expansion of Eq.~\eqref{G_analytic_t0} using symbolic computation tools such as \textsc{Mathematica}. As we show in the next section, the WCE/SCE series obtained from the Taylor expansions indeed match those derived from the diagrammatic expansion. This agreement is not coincidental, but rather follows directly from the uniqueness of the analytic function. To see this, we analytically continue \(U \in \mathbb{R}\) to the complex plane for any fixed \(t \neq 0 \in \mathbb{R}\). 
Since Eq.~\eqref{G_analytic_t0} defines an algebraic function in the variables \((t,U)\), for any fixed $t$, this analytic continuation is well-defined and the resulting Green's function $G_{ij}(i\omega_n, U)$ can be verified to be analytic at \(U=0\). Therefore, it admits a {\em unique} Taylor expansion around \(U=0\), implying that the WCE obtained from the Taylor expansion must coincide with the one derived from the diagrammatic expansion. The similar argument applies to the SCE by analytically continuing in \(t\).

One may wonder why the WCE series is \emph{unique}, given that different choices of the bare Green’s function \(G_0\) could, in principle, lead to different diagrammatic expansions. This is because the WCE coefficients $G^{(m)}(t,i\omega_n)$ defined in Eq.~\eqref{wce_taylor} is {\em independent} of $U$. For the half-filling case where $\mu=U/2$, this actually requires the bare Green's function to be the HF Green’s function. In fact, as mentioned in \cite{ferrero2023diagrammatic}, the HF \(G_0\) is also the optimal bare propagator to construct a locally convergent WCE series for the half filling case. 

\subsection{Two-dimensonal Hubbard model}
\label{app:sce_hubbard}
We need to use diagrammatic method to derive the WCE/SCE of 2D Hubbard model.  In the momentum-frequency space, up to the second-order, the WCE series is given by 
\begin{equation}\label{eqn:Sigma}
\begin{aligned}
G(\bm k,i\omega_n)
=G_0(\bm k,i\omega_n)+ G_0(\bm k,i\omega_n)\Sigma^{2ndB}(\bm k,i\omega_n)G_0(\bm k,i\omega_n)
\end{aligned}
\end{equation}
where $G_0(\bm k,i\omega_n)$ is the HF Green's function, $\Sigma^{2ndB}=\Sigma^{2ndB}[G_0](\bm k,i\omega_n)$ denotes the bare second-order contribution to the self-energy, corresponding to Feynman diagram
\begin{fmffile}{2ndB}
\raisebox{-0.1cm}{
    \resizebox{0.03\textwidth}{!}{ 
    \begin{fmfgraph*}(40,30)
      \fmftop{v1,v3}
      \fmfbottom{v2,v4}
      \fmf{photon}{v1,v2}
      \fmf{photon}{v3,v4}
      \fmf{plain_arrow,left=0.4,tension=0.4}{v1,v3,v1}
      \fmf{fermion}{v2,v4}
    \end{fmfgraph*}
    } 
}
\end{fmffile}
. This choice of \(G_0(\boldsymbol{k}, i\omega_n)\) ensures the optimal convergence rate of the WCE series~\cite{ferrero2023diagrammatic} for the half-filling case. 
As illustrated in the Hubbard dimer example, the corresponding chemical-potential shift guarantees that \(G_0(\boldsymbol{k}, i\omega_n)\), and hence all expansion coefficients \(G^{(m)}(\boldsymbol{k}, i\omega_n)\) remain independent of \(U\). Another consequence of this is that, at second order, only the diagram mentioned above contributes while all other diagrams vanish.

The SCE for the Matsubara Green's function of the Hubbard model at half-filling is given (explicit up to second order) by \cite{metzner1991linked}:
\begin{equation}\label{SCE_2d_hubbard}
\begin{aligned}
G_{ij}(i\omega_n) &= G^{(0)}_{ij}(i\omega_n) + G^{(1)}_{ij}(i\omega_n) + G^{(2)}_{ij}(i\omega_n) + \cdots \\
&= \frac{i\omega_n}{(i\omega_n)^2 - \frac{U^2}{4}} 
- \frac{(i\omega_n)^2}{\left[(i\omega_n)^2 - \frac{U^2}{4}\right]^2} t_{ij} \\
&\quad + \frac{(i\omega_n)^3}{\left[(i\omega_n)^2 - \frac{U^2}{4}\right]^3} 
\left( \sum_k t_{ik} t_{kj} - \sum_k t_{ik} t_{ki} \delta_{ij} \right) \quad + \frac{3}{4} \frac{i\omega_n U^2}{\left[(i\omega_n)^2 - \frac{U^2}{4}\right]^3} 
\sum_k t_{ik} t_{ki} \delta_{ij} + \cdots
\end{aligned}
\end{equation}
\paragraph{Reformulating SCE into a series of $U$} The SCE series we obtained for the Hubbard dimer and the general Hubbard model is expressed in the hopping amplitude \(t\) (or \(|t_{ij}|\) in the lattice case), and is therefore valid in the limit \(t \to 0\) for any fixed interaction strength \(U\). 
To combine this with the WCE and construct the two-point Padé expansion, the SCE must be reformulated as a series in the variable \(U\). 
In this work, we achieve this reformulation via a rescaling transformation. 
Specifically, if we aim to obtain an SCE for a Hubbard model with fixed \(t = 1.0\) and treating \(U\) as the expansion variable, then the Matsubara Green’s function satisfies
\begin{align}\label{scaling_transform}
G_{ij}(t = 1.0, U, i\omega_n)
= \frac{1}{U}\,
G_{ij}\!\left(t = \frac{1}{U},\, U' = 1.0,\, \frac{i\omega_n}{U}\right).
\end{align}
As \(U \to +\infty\) and \(t = 1/U \to 0\), the right-hand side of Eq.~\eqref{scaling_transform} can be approximated using the SCE in Eqs.~\eqref{sce_taylor} or~\eqref{SCE_2d_hubbard}, depending on whether the model is Hubbard dimer. 
Let \(G^{\mathrm{app},M}\) denote the \(M\)-th order approximation of $G$ with respect to $t$. 
Using \(G^{\mathrm{app},M}\), we evaluate \(G(t=1.0, U, i\omega_n)\) and its derivatives with respect to $U$ at a chosen expansion point \(U=U_0\) numerically via finite-difference by following the right-hand side of Eq.~\eqref{scaling_transform}. 
This yields an approximate Taylor expansion of \(G_{ij}(t=1.0, U, i\omega_n)\) about \(U = U_0\), whose accuracy systematically improves as \(U_0,\,M \to +\infty\). 
By combining this SCE-based Taylor expansion with the WCE, we then employ the Padé–Taylor approach to construct a global approximation to \(G(t=1.0, U, i\omega_n)\). In practice, we normally choose $10\leq U_0\leq 20$, where the specific $U_0$ value can be adjusted to ensure the Pad\'e interpolation result is numerically stable.

%
%
\section{Analytic continuation for solvable models}
\begin{figure}[t]
\centering
\includegraphics[width=16.5cm]{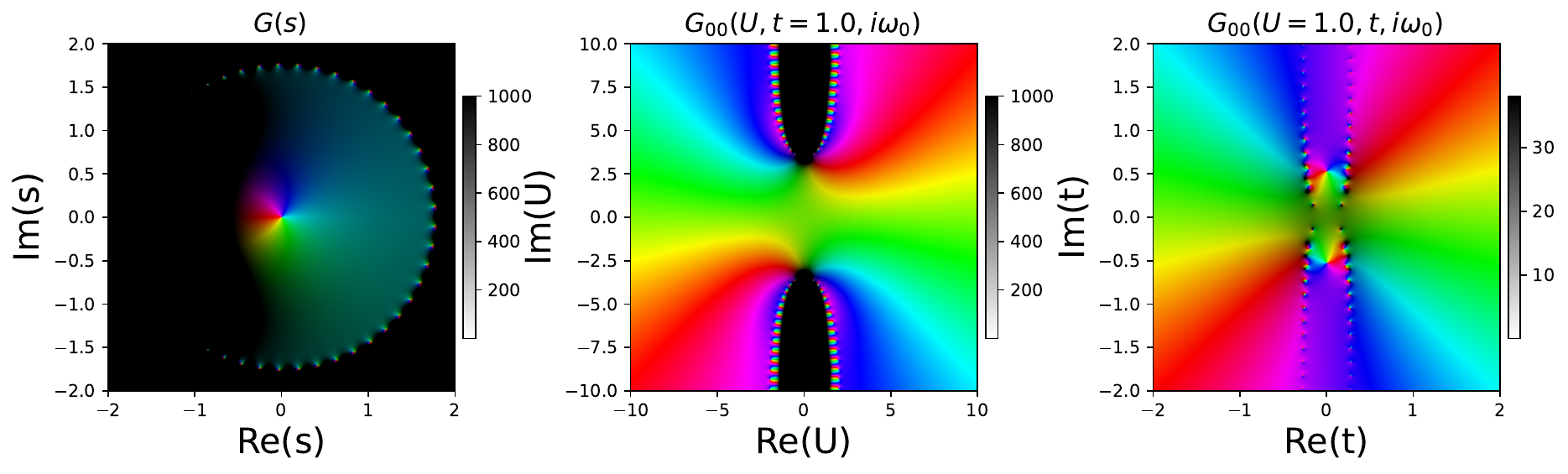}
\caption{Analytic continuation (AC) of the correlation functions for exactly solvable models. In each panel, grayness indicates the modulus of
the function, while the hue represents the complex argument.
\textbf{(Left)} AC of the correlation function \( G(s) \) for the zero-dimensional \( \phi^4 \) model. 
The plot is obtained from an approximated \( G(s) \) based on the truncated SCE series and therefore only accurately represents the function’s behavior in the vicinity of \( s = 0 \).
\textbf{(Middle)} AC of the exact diagonal Matsubara Green’s function \( G_{00}(U, t = 1.0, i\omega_n) \) of the Hubbard dimer. 
The displayed results correspond to the Matsubara frequency \( i\omega_0 = i\pi / \beta \) with \( \beta = 20 \); AC results for other frequencies exhibit the same features. 
\textbf{(Right)} AC of the exact diagonal Matsubara Green’s function \( G_{00}(U = 1.0, t, i\omega_n) \) of the Hubbard dimer, also shown at \(i\omega_0 = i\pi / \beta \) and \( \beta = 20 \), with similar behavior observed at other frequencies. 
}
\label{fig:AC_solvable_models}
\end{figure}
In this section, we perform analytic continuation (AC) of the correlation functions for the two exactly solvable models considered in this work: the zero-dimensional \( \phi^4 \) model and the Hubbard dimer. 
The AC results clearly demonstrate that the pole structures of the correlation are consistent with the illustrative figure presented in the main text.
\paragraph{Zero-dimensional $\phi^4$ model} The WCE for the zero-dimensional \( \phi^4 \) model is known to be an asymptotic series in \( g \), and \( g = 0 \) is a branch point if one directly performs analytic continuation in \( g \) for the analytic solution, as shown in Ref.~\cite{brown2015two}. The convergence of the two-point Padé expansion originates from the analyticity of the SCE at \( s = 0 \), where \( s = 1/\tilde{g} = 1/\sqrt{g} \). 
After the conformal transformation \(f(z) = 1/\sqrt{z} \), the SCE series Eq.~\eqref{0d_sce} can be re-expressed in terms of $s$ as:
\begin{align}
G(s) &= \left[\frac{2\sqrt{6}\,\Gamma\left(\frac{3}{4}\right)}{\Gamma\left(\frac{1}{4}\right)}\right]s
+ 12\left[\left(\frac{\Gamma\left(\frac{3}{4}\right)}{\Gamma\left(\frac{1}{4}\right)}\right)^2 - \frac{\Gamma\left(\frac{5}{4}\right)}{\Gamma\left(\frac{1}{4}\right)}\right]s^2+\cdots
\label{0d_sce_s}
\end{align}
which can be analytically continued to the complex \( s \) plane and yielding a result different from the AC of the analytic solution after replacing $g$ with $1/s^2$. The approximated complex-valued function \( G(s) \) is shown in the left panel of Fig.~\ref{fig:AC_solvable_models}. 
To generate this plot, we truncated the expansion in Eq.~\eqref{0d_sce_s} at the 50th order. 
Due to this truncation, the plot only accurately represents \( G(s) \) in the vicinity of \( s = 0 \). 
From the figure, we observe that \( G(s) \) possesses an analytic germ at \( s = 0 \). 
Hence, by successive analytic continuation, the original function along the positive real axis can be recovered. 
Numerically, this continuation is implemented using the two-point Padé expansion by also including the series expansion at $s=+\infty$ (WCE in term of $g$). The final AC results evaluated along the real axis are presented in the main text. Note that in the extended $s$-plane, the $s=\infty$ point is {\em not} a analytic germ due to the asymptotic nature of the WCE series in terms of $g$. As a result, the two-point Pad\'e expansion does not interpolate between two analytic germs as displayed in the main text, but the convergence is still possible as explained in \cite{buslaev2013convergence}. For the $\phi^4$ example, we also see that numerically the approximation works well. 

\paragraph{Hubbard dimer} In contrast to the \( \phi^4 \) model, the WCE of the Matsubara Green’s function $G_{ij}(U,t,i\omega_n)$ for the Hubbard dimer is locally convergent, i.e., it possesses an analytic germ at $U=0$ in the complex-$U$ plane for any fixed $t>0$ and $i\omega_n$. 
This property is also expected to hold for Hubbard model in a finite lattice and at finite temperature, which serves as the rationale behind the development of diagrammatic QMC methods~\cite{ferrero2023diagrammatic} for computing high-order expansion series. 
For the Hubbard dimer, the pole structure can be directly visualized through the analytic continuation of the exact solution~\eqref{G_analytic_t0}. 
For \( t = 1.0, \beta =20\), the resulting complex-valued function is shown in the middle panel of Fig.~\ref{fig:AC_solvable_models}. 
From the figure, we clearly observe an analytic germ at \( U = 0 \), with all singularities located away from the positive real axis. 
The analytic continuation with respect to \( t \) is shown in the right panel of Fig.~\ref{fig:AC_solvable_models}. 
Similarly, the singularities remain away from the positive real axis; however, the poles appear much closer to \( t = 0 \), indicating that the SCE in \( t \) has a much smaller convergence radius than the WCE.

\section{Additional numerical results}
\begin{figure}[t]
\centering
\includegraphics[width=12cm]{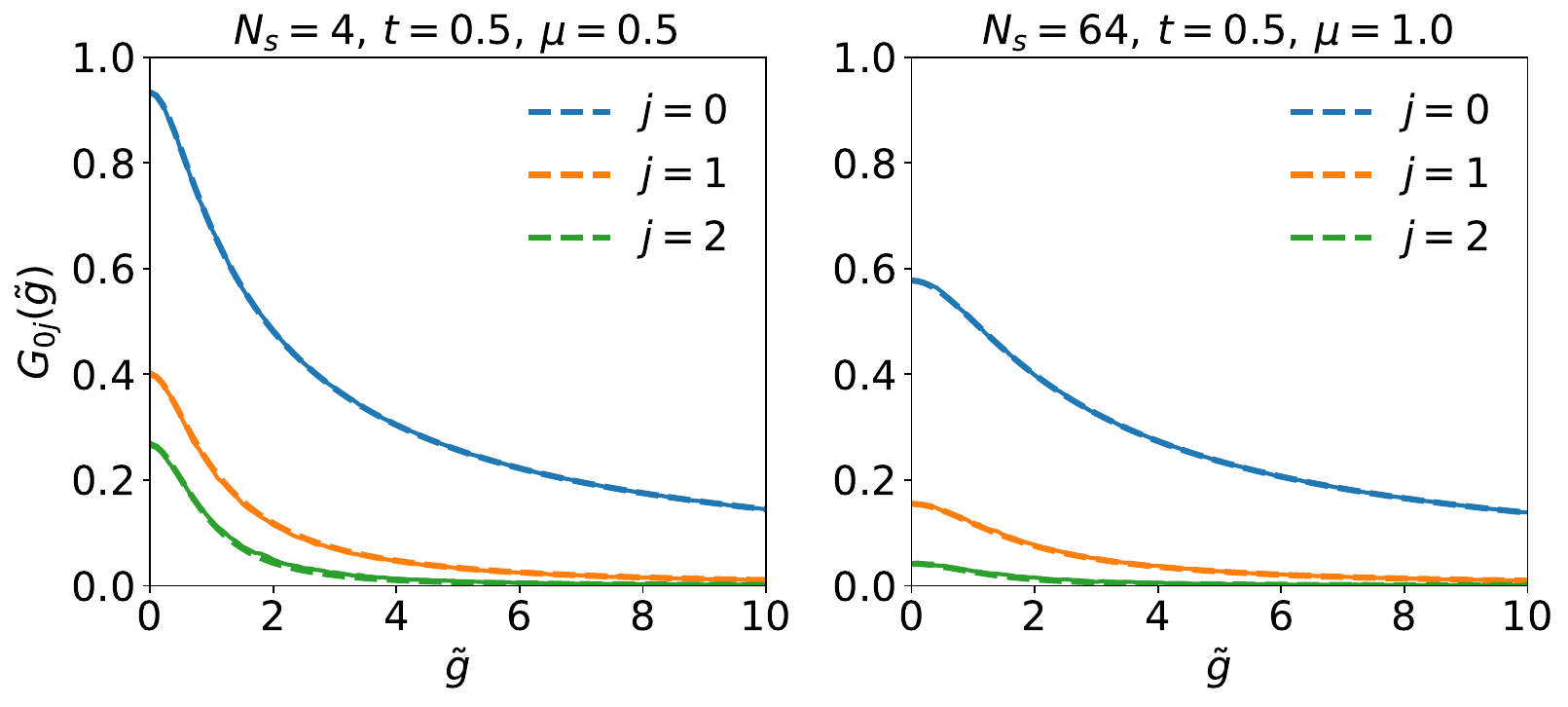}
\caption{Off-diagonal correlation functions \( G_{0j}(\tilde{g}) \) of the one-dimensional lattice \( \phi^4 \) model. 
In each panel, the solid lines represent the ground-truth results obtained from Langevin Monte Carlo simulations, 
while the dashed lines show the corresponding approximations from the two-point Padé scheme (Padé-\(w_6\)-\(s_3\)).
}
\label{fig:1d_phi4_offd}
\end{figure}
\begin{figure}
\centering
\includegraphics[width=12cm]{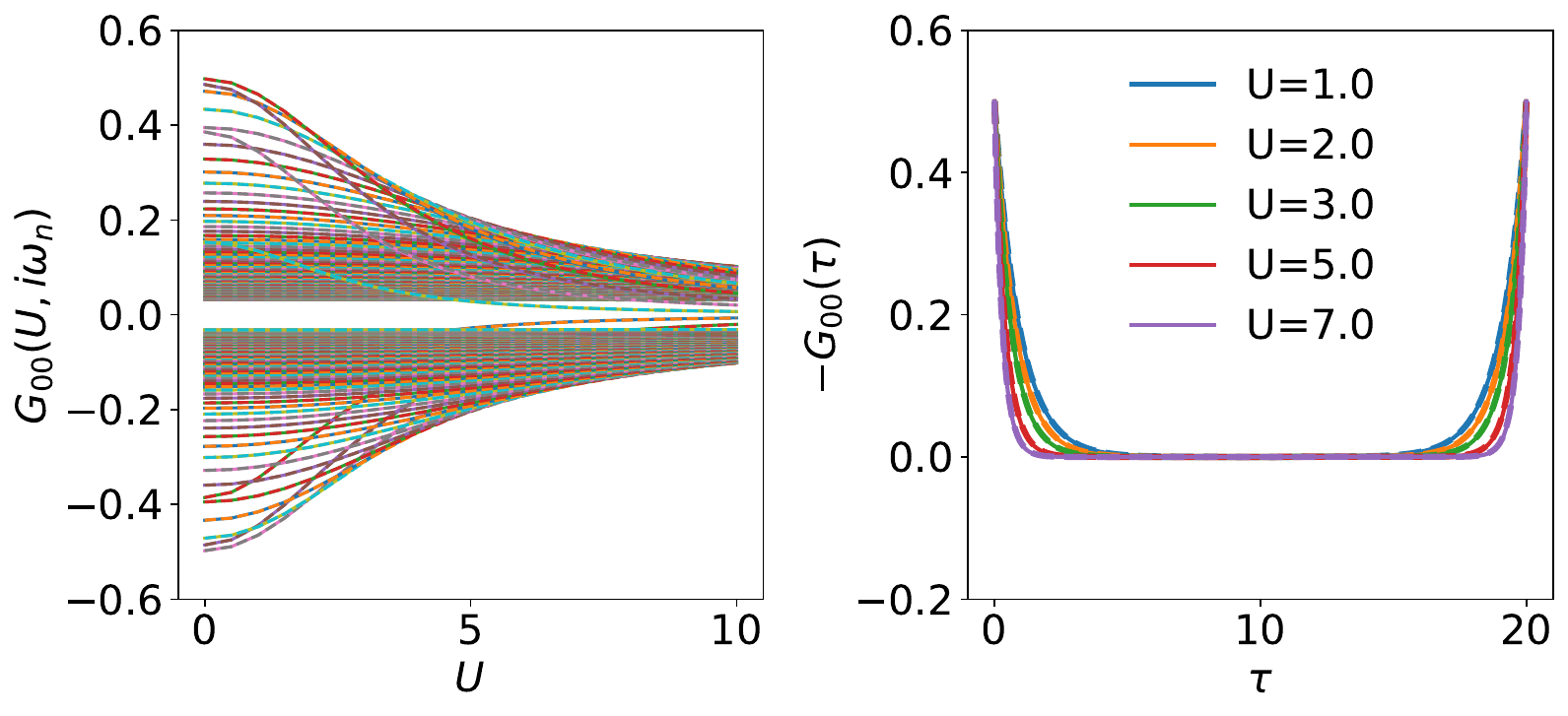}
\caption{\textbf{(Left)} Approximation of the imaginary part of the Hubbard dimer Green’s function, 
\(\mathrm{Im}[G_{00}(U, i\omega_n)]\), using the Padé–Taylor scheme (Padé-\(w_4\)–\(s_3\)). 
The curves show interpolation results for all Matsubara frequencies \(i\omega_n\) 
\textbf{(Right)} The corresponding Green’s function \(G_{00}(U, \tau)\) in the imaginary-time domain. In both figures, solid lines represent the analytic solution and the dashed lines correspond to the Padé approximation. All the results are obtained for $t=1.0, \beta=20$.
}
\label{fig:hubbard_dimer_interpolation}
\end{figure}

\begin{figure}
\centering
\includegraphics[width=15cm]{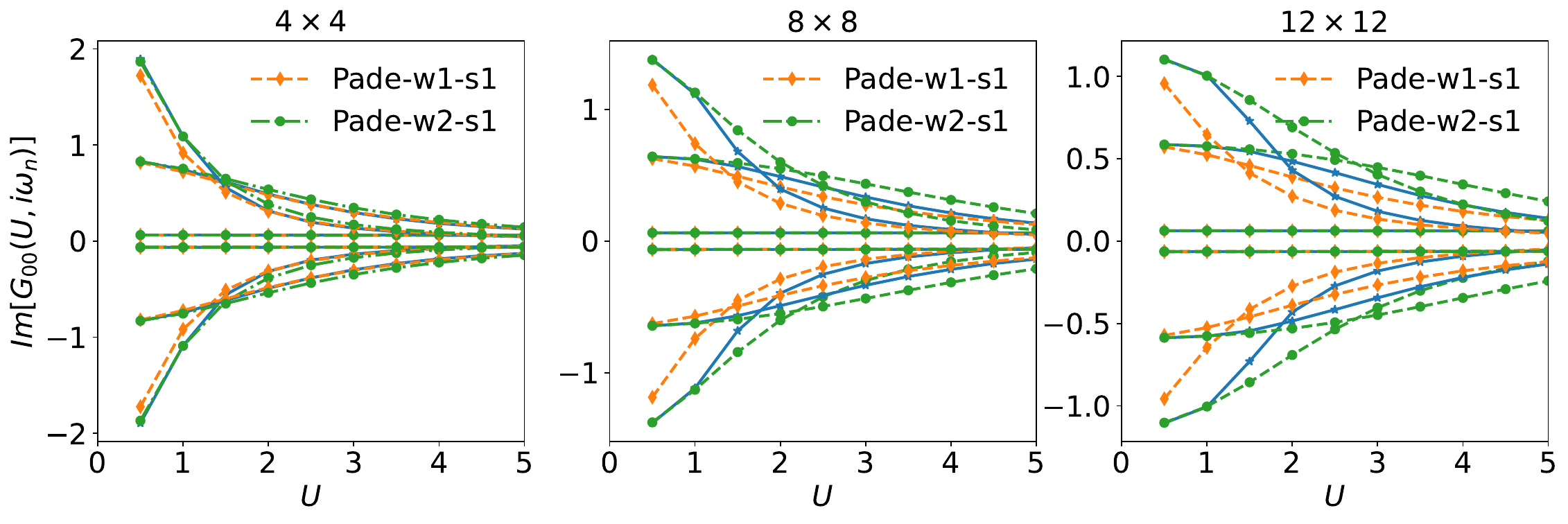}
\caption{Approximation of the Matsubara Green’s function for the 2D Hubbard model with different square lattice sizes, obtained using two Padé–Taylor schemes. 
The blue curve with star markers indicates the AFQMC benchmarks. 
Each panel compares the approximation results at selected Matsubara frequencies (from top to down): 
\(n =-1,-2,-50, 49, 1, 0\), where $\omega_n=(2n+1)\pi/\beta$.
All calculations are performed with \( t = 1.0 \) and \( \beta = 20 \).
}
\label{fig:2d_hubbard_interpolation}
\end{figure}
In this section, we present additional numerical results that further support the main arguments of the paper. 
Figure~\ref{fig:1d_phi4_offd} shows that the two-point Padé expansion also provides an accurate approximation for the off-diagonal elements of the correlation function \( G_{ij} = \langle \phi_i \phi_j \rangle \) in the lattice \( \phi^4 \) field model. 
For the Hubbard dimer, Fig.~\ref{fig:hubbard_dimer_interpolation} illustrates how the two-point interpolation with respect to \( U \) accurately approximates the Matsubara Green’s function \( G_{00}(U, i\omega_n) \) at different Matsubara frequencies (left panel) and how the corresponding approximation behaves in the imaginary-time domain (right panel). For the two-dimensional Hubbard model, Fig.~\ref{fig:2d_hubbard_interpolation} demonstrates how the Padé approximation works as the system size increases. 
From the figure, we first observe that the approximation systematically improves as the two-point Padé expansion order increases. 
Nevertheless, higher-order calculations are still needed to confirm whether this trend persists like the dimer case and eventually lead to a convergent, global approximation to the Green's function. 
On the other hand, as the system size increases from \(4\times4\) to \(8\times8\) and \(12\times12\), the comparison with AFQMC benchmarks—particularly at low Matsubara frequencies—clearly indicate that higher-order expansions are required to construct more accurate approximations of the Green’s function. 
This is evident from the change in the underlying function’s curvature around \( U = 1.0 \), where it transitions from concave down to concave up, requiring at least third-order WCE terms (if nonzero) to capture this behavior.

\bibliographystyle{plain}
\bibliography{Derivation}